\documentclass[default,iicol]{sn-jnl}

\jyear{2022}%

\theoremstyle{thmstyleone}%
\newtheorem{theorem}{Theorem}

\theoremstyle{thmstyletwo}%
\newtheorem{example}{Example}%

\theoremstyle{thmstylethree}%
\newtheorem{definition}{Definition}%

\begin{document}

\title[The Quantum-House Effect]{The Quantum-House Effect and Its Demonstration on~SpinQ~Gemini}

\author*{\fnm{Tamás} \sur{Varga}}\email{tvarga@q-edu-lab.com}

\affil{\orgname{q-edu-lab.com}, \orgaddress{\state{Zurich}, \country{Switzerland}}}

\abstract{We introduce the quantum-house effect, a non-local quantum phenomenon which goes against classical intuition. We show how the effect can be achieved with any bipartite quantum state where neither subsystem is in a pure state. Besides its theoretical description, the quantum-house effect is also demonstrated on SpinQ Gemini, a 2-qubit liquid-state NMR desktop quantum computer.}

\keywords{quantum-house effect, quantum nonlocality, SpinQ Gemini, nuclear magnetic resonance (NMR) desktop quantum computer}

\maketitle

\section{Introduction}\label{sec1}

Quantum entanglement was discovered nearly a century ago \cite{einstein}, and has become a signature effect of quantum mechanics. Schrödinger called entanglement \textit{the} characteristic trait of quantum mechanics, which alone embodies the difference between quantumness and classicality \cite{schrodinger}.

In the past decades, entanglement has also turned out to be a key resource in quantum information processing \cite{nielsen,schumacher}. In particular, several authors pointed out that entangled quantum states play an essential role in achieving exponential speed-up in certain quantum-computing algorithms \cite{ekert}, and others hinted that in the absence of entanglement one should not talk about "true" quantum computation, but rather a simulation thereof \cite{braunstein}.

In this paper,\footnote{An extended version of this paper, covering a wider scope, can be found in \cite{varga}.} after explaining basic concepts of quantum information, we go beyond entanglement and introduce the quantum-house effect, a non-local quantum phenomenon which can be exhibited even with bipartite product states. This indicates that with respect to non-locality, quantum systems can behave in a counter-intuitive way already without entanglement.

The quantum-house effect can be considered as an extension of locally non-effective unitary operations, first proposed in \cite{fu} and further investigated in \cite{datta}.\footnote{The present work came about independently of \cite{fu}. The core idea we arrived at is basically the same, but our original motivation was educational, to explore quantum effects that can be demonstrated on SpinQ Gemini.}

\begin{figure}[t]
\centering
\includegraphics[width=0.7\columnwidth]{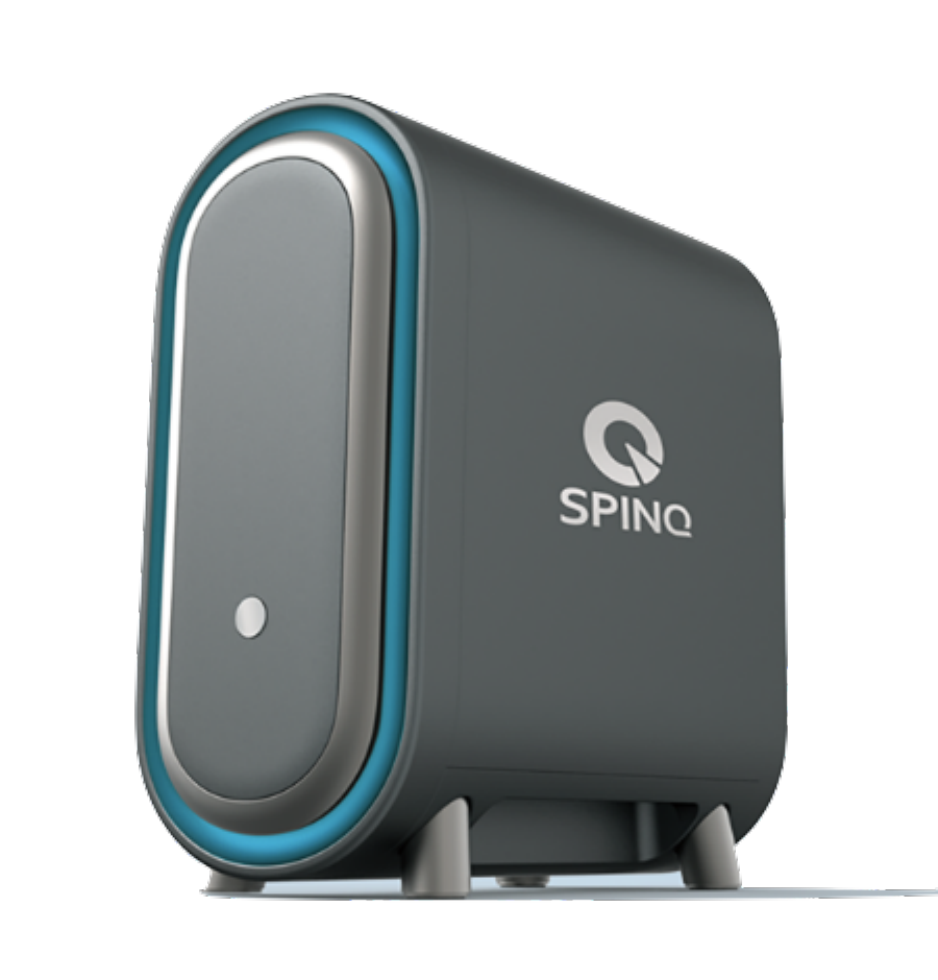}
\caption{SpinQ Gemini, a 2-qubit NMR desktop quantum computer. Source: \url{https://www.spinq.cn/}}\label{figgemini}
\end{figure}

The paper is organized as follows. In Section~\ref{secedu}, we briefly review quantum states and entanglement. Then, Sections~\ref{sec2} and \ref{sec4} present theoretical results, while in Section~\ref{sec3} we demonstrate the quantum-house effect using SpinQ Gemini, a 2-qubit liquid-state nuclear magnetic resonance (NMR) desktop quantum computer \cite{hou}, shown in Fig.~\ref{figgemini}. Finally, Section~\ref{sec5} discusses our findings and suggests a principle to characterize where quantumness departs from classicality.

\section{Preliminaries}\label{secedu}

In this section, we briefly review those concepts of quantum information that are essential for the understanding of the rest of the paper. For a comprehensive introduction to the same, the reader is referred to \cite{nielsen} or \cite{schumacher}. The necessary mathematics is concisely summarized in \cite{woody}.

\subsection{Superposition and the qubit}\label{eduqubit}

According to quantum theory, isolated physical systems may be in \textit{superposition} of their reliably distinguishable states.

As an example, let's take a physical system that represents a bit in a computer. If the same system was isolated from its environment, then in principle it could represent a \textit{qubit}, which can be in superposition where it is partly $\vert 0\rangle$ and partly $\vert 1\rangle$, thus in a sense both $\vert 0\rangle$ and $\vert 1\rangle$ at the same time.\footnote{In quantum mechanics, it is customary to write the state of an isolated physical system between the symbols $\vert$ and $\rangle$, following the convention called "bra-ket notation".}

Physically, a (qu)bit can be realized by two internal energy levels of an atom, $E_0$ and $E_1$, playing the role of $\vert 0\rangle$ and $\vert 1\rangle$ (here, superposition could be interpreted as having both energies $E_0$ and $E_1$ at the same time), or by other physical systems like the spin of an individual electron or photon.

Analogously to the role the bit plays in classical information processing, the qubit can be considered as the basic unit of quantum information processing.

\subsection{State vectors in Hilbert space}\label{edustatevec}

The quantum state of an isolated physical system can be completely described by the \textit{state-vector formalism}.

The idea is that mathematically, every possible state $\vert\psi\rangle$ of the system is a unit vector in an associated $N$-dimensional complex Hilbert space, the so-called "state space", where $N$ depends on the system in question.\footnote{In this paper, only finite-dimensional Hilbert spaces are considered. Although it can be challenging to find the appropriate Hilbert space for a particular physical system, quantum theory postulates such a state space always exists.} Two physical states are reliably distinguishable if and only if they are represented by orthogonal vectors (i.e. whose inner product is zero) in the state space.

\begin{example}\label{exqubitatom}
If we use the atomic energy levels $E_0$ and $E_1$ to realize a qubit, the associated mathematical state vectors $\vert E_0\rangle$ and $\vert E_1\rangle$ must be orthogonal, since $E_0$ and $E_1$ can be reliably distinguished by measuring the internal energy of the atom.
\end{example}

As our topic is quantum information, in the following we'll focus on abstract systems (and associated Hilbert spaces), especially those made up of qubits, rather than concrete physical systems. It is analogous to focusing on (abstract) bits when dealing with classical information.

\begin{example}\label{exqubitabs}
The (abstract) qubit has two classical states, $\vert 0\rangle$ and $\vert 1\rangle$. Ideally, classical states can be distinguished reliably, so $\vert 0\rangle$ and $\vert 1\rangle$ must be orthogonal unit vectors in the qubit's state space.\footnote{If it is more comfortable having a concrete physical system in mind, identify $\vert 0\rangle$ with $\vert E_0\rangle$ and $\vert 1\rangle$ with $\vert E_1\rangle$.}
\end{example}

\subsection{Basis states, amplitudes, inner product}\label{edubasisvec}

Let $B=\left\{\vert b_0\rangle,\vert b_1\rangle,\ldots,\vert b_{N-1}\rangle\right\}$ be an orthonormal basis (i.e. pairwise orthogonal unit vectors) of the state space associated with a system. Then, any quantum state $\vert\psi\rangle$ can be written as a unique linear combination of the \textit{basis states} in $B$:
\begin{equation}\label{eqampl}
\vert\psi\rangle=a_0\vert b_0\rangle+a_1\vert b_1\rangle+\ldots+a_{N-1}\vert b_{N-1}\rangle
\end{equation}

The complex coefficients $a_0,a_1,\ldots,a_{N-1}$ are called \textit{amplitudes}. Intuitively, each amplitude $a_i$ indicates "how much" the basis state $\vert b_i\rangle$ participates in $\vert\psi\rangle$. Having at least two non-zero amplitudes means that $\vert\psi\rangle$ is a superposition state of more than one basis state.\footnote{Using complex amplitudes to express superposition is a clever mathematical trick that makes quantum states amenable to linear algebra treatment. That said, it is not obvious at all why all sorts of quantum physical systems we know of admit a description in terms of \textit{complex} Hilbert spaces.}

Given the orthonormal basis $B$ (with indexed elements), we can also identify $\vert\psi\rangle$ with the following column vector, i.e. $N\times 1$ matrix:
\begin{align}\label{eqketcol}
\vert\psi\rangle &= \begin{bmatrix}
          a_0 \\
          a_1 \\
          \vdots \\
          a_{N-1}
        \end{bmatrix}
\end{align}

It is perhaps easier to think of a quantum state like this, even if we know that the column vector depends on the chosen basis $B$. This form, as we'll see next, is also convenient when calculating the inner product of two state vectors. In the bra-ket notation, $\vert\psi\rangle$ is called a \textit{ket}, while $\langle\psi\vert$ a \textit{bra}. Given $B$, the latter can be identified with a row vector, i.e. $1\times N$ matrix, as follows:
\begin{equation}\label{eqrow}
\langle\psi\vert=\begin{bmatrix}a_0^* & a_1^* & \ldots & a_{N-1}^*\end{bmatrix}
\end{equation}

Here, $a_i^*$ is the complex conjugate of $a_i$. In matrix language, $\langle\psi\vert$ is the conjugate transpose of $\vert\psi\rangle$. With that, the inner product of $\vert\psi\rangle$ and $\vert\phi\rangle$, denoted by $\langle\psi\vert\phi\rangle$, can be calculated by matrix multiplication, multiplying the row vector $\langle\psi\vert$ with the column vector $\vert\phi\rangle$. The requirement that any quantum state $\vert\psi\rangle$ must be a unit vector means $\langle\psi\vert\psi\rangle=\sum_{i=0}^{N-1}\vert a_i\vert^2=1$ must hold.\footnote{Conversely, in the context of quantum information, we can assume that every unit vector $\vert\psi\rangle$ represents a legitimate quantum state (of an abstract system, e.g. a qubit).}

\begin{example}\label{excompbasis}
The set $B=\left\{\vert 0\rangle,\vert 1\rangle\right\}$ is called the \textit{computational basis} of the qubit. Any qubit state $\vert\psi\rangle$ can be written as $\vert\psi\rangle=a_0\vert 0\rangle+a_1\vert 1\rangle$, with some unique amplitudes $a_0$ and $a_1$. Therefore, in terms of the computational basis, $\vert 0\rangle$, $\vert 1\rangle$ and $\vert\psi\rangle$ can be identified with the column vectors $\begin{bmatrix}1\\0\end{bmatrix}$, $\begin{bmatrix}0\\1\end{bmatrix}$ and $\begin{bmatrix}a_0\\a_1\end{bmatrix}$, respectively. As an exercise, we can now calculate that the inner product of $\vert 0\rangle$ and $\vert 1\rangle$ is zero, thus they are indeed orthogonal: $\langle 0\vert 1\rangle=\begin{bmatrix}1 & 0\end{bmatrix}\begin{bmatrix}0\\1\end{bmatrix}=1\cdot 0+0\cdot 1=0$.
\end{example}

Finally, a peculiarity of the state-vector formalism is that $\vert\psi\rangle$ and $c\vert\psi\rangle$ represent the very same physical state, where $c$ is any complex number with $\lvert c\rvert=1$. So for example $\vert b_i\rangle$ and $-\vert b_i\rangle$ are equivalent, they mean the same thing.

\subsection{Manipulating state vectors}\label{edumanipulate}

The basic operations to manipulate the state vector are unitary transformations and measurements in a basis. Any other operations allowed on a quantum system can be thought of as arising from these basic ones \cite{vedral}.

\textbf{Unitary transformation.} A unitary transformation is a linear operation $U$ that brings unit vectors to unit vectors. Given the orthonormal basis $B$, $U$ can be identified with an $N\times N$ unitary matrix with which the state column vector is to be multiplied to get the new state vector. When acting on qubits, a unitary transformation is also called a \textit{quantum gate}.\footnote{Again, in the context of quantum information, we can assume that every unitary transformation $U$ represents a legitimate state manipulation (of an abstract system, e.g. a qubit).}

Let's see two examples, using the computational basis:

\begin{example}\label{exunitaryx}
The 1-qubit Pauli-$X$ gate is identified with the $X=\begin{bmatrix}0 & 1\\1 & 0\end{bmatrix}$ unitary matrix. The gate's effect can be calculated by matrix multiplication, e.g. it negates the classical states: $X\vert 0\rangle=\begin{bmatrix}0 & 1\\1 & 0\end{bmatrix}\begin{bmatrix}1\\0\end{bmatrix}=\begin{bmatrix}0\\1\end{bmatrix}=\vert 1\rangle$ and $X\vert 1\rangle=\begin{bmatrix}0 & 1\\1 & 0\end{bmatrix}\begin{bmatrix}0\\1\end{bmatrix}=\begin{bmatrix}1\\0\end{bmatrix}=\vert 0\rangle$. In general, it swaps the amplitudes of $\vert 0\rangle$ and $\vert 1\rangle$, that is: $X\left(a_0\vert 0\rangle+a_1\vert 1\rangle\right)=\begin{bmatrix}0 & 1\\1 & 0\end{bmatrix}\begin{bmatrix}a_0\\a_1\end{bmatrix}=\begin{bmatrix}a_1\\a_0\end{bmatrix}=a_1\vert 0\rangle+a_0\vert 1\rangle$.
\end{example}

\begin{example}\label{exunitaryh}
The 1-qubit Hadamard gate is identified with the $H=\frac{1}{\sqrt{2}}\begin{bmatrix}1 & 1\\1 & -1\end{bmatrix}$ unitary matrix. This gate can be used to prepare superposition of classical states: $H\vert 0\rangle=\frac{1}{\sqrt{2}}\begin{bmatrix}1 & 1\\1 & -1\end{bmatrix}\begin{bmatrix}1\\0\end{bmatrix}=\frac{1}{\sqrt{2}}\begin{bmatrix}1\\1\end{bmatrix}=\frac{1}{\sqrt{2}}\vert 0\rangle+\frac{1}{\sqrt{2}}\vert 1\rangle$ and $H\vert 1\rangle=\frac{1}{\sqrt{2}}\begin{bmatrix}1 & 1\\1 & -1\end{bmatrix}\begin{bmatrix}0\\1\end{bmatrix}=\frac{1}{\sqrt{2}}\begin{bmatrix}1\\-1\end{bmatrix}=\frac{1}{\sqrt{2}}\vert 0\rangle-\frac{1}{\sqrt{2}}\vert 1\rangle$. So unlike the Pauli-$X$ gate, the Hadamard gate brings classical states to non-classical states, and as such it is "quantum-native", with no classical counterpart.
\end{example}

\textbf{Measurement in a basis.} Measuring the system in an orthonormal basis $B$ may randomly yield $N$ different numerical results, each associated with finding the system in a corresponding basis state of $B$. The probability of each outcome is given by the \textit{Born rule}: if the system's state is $\vert\psi\rangle=a_0\vert b_0\rangle+a_1\vert b_1\rangle+\ldots+a_{N-1}\vert b_{N-1}\rangle$ right before the measurement, it will \textit{collapse} with probability $\vert a_i\vert^2$ into the basis state $\vert b_i\rangle$, for every $0\le i<N$, due to the act of measuring.\footnote{Note that $a_i=\langle b_i\vert\psi\rangle$, and $\vert a_i\vert^2=a_ia_i^*=\langle b_i\vert\psi\rangle\langle\psi\vert b_i\rangle$.}

\begin{example}\label{exmeasureatom}
Let's use the atomic energy levels $E_0$ and $E_1$ to realize a qubit. Then, measuring the internal energy of the atom is a measurement in the $B=\left\{\vert E_0\rangle,\vert E_1\rangle\right\}$ basis. If the state is e.g. $\vert\psi\rangle=0.6\vert E_0\rangle+0.8\vert E_1\rangle$ right before the measurement, then with probability $0.36$ we'll get the numerical result $E_0$ and the new (collapsed) state will be $\vert E_0\rangle$ right after the measurement. Or, with probability $0.64$ we'll get $E_1$ with the corresponding new state being $\vert E_1\rangle$. Since getting $E_0$ or $E_1$ are the only possibilities, $\vert 0.6\vert^2+\vert 0.8\vert^2=1$ holds, explaining why $\vert\psi\rangle$ must be a unit vector.
\end{example}

\begin{example}\label{exmeasurequbit}
Let a qubit's state be $\vert\psi\rangle=a_0\vert 0\rangle+a_1\vert 1\rangle$. Measuring it in the computational basis, we'll get $0$ with probability $\vert a_0\vert^2$ and $1$ with probability $\vert a_1\vert^2$. The corresponding post-measurement (collapsed) state will be $\vert 0\rangle$ and $\vert 1\rangle$, respectively. Since these are the only possibilities, $\vert a_0\vert^2+\vert a_1\vert^2=1$ must hold. (The numerical results $0$ and $1$ may be chosen freely here, as the qubit is an abstract system.)
\end{example}

Unlike classical physics, in quantum mechanics we cannot "see" the state of a system. The only way to observe is to make measurements, i.e. operations that yield a numerical result.\footnote{In this paper, only measurements in a basis are considered.} The catch is that the result is inherently random, and the state is disturbed via collapse. All that can be known in advance is the probability distribution of the possible outcomes. Accordingly, the "quantum state" can be thought of as a mathematical entity (such as a vector $\vert\psi\rangle$ in a Hilbert space) that lets us calculate this probability distribution for \textit{any} conceivable measurement on the system \cite{mermin,hardy}. Or, equivalently, as a mathematical entity that captures all that an experimenter can (statistically) find out about the system if she is given a large sample of identically prepared instances.

\subsection{Bipartite systems and entanglement}\label{edubipartite}

Let $A$ and $B$ be two distinct physical systems. The composite system $AB$ (i.e. $A$ and $B$ together) is called a \textit{bipartite system}. According to quantum theory, the state space associated with $AB$ is the tensor product of the two underlying Hilbert spaces associated with subsystems $A$ and $B$, and the possible states of $AB$ can be described as follows.

\textbf{Product state.} If $A$ and $B$ are prepared independently, in state $\vert\alpha\rangle$ and $\vert\beta\rangle$, respectively, the state vector for $AB$ is $\vert\psi\rangle=\vert\alpha\rangle\otimes\vert\beta\rangle$, i.e. the tensor product of $\vert\alpha\rangle$ and $\vert\beta\rangle$.

\begin{example}\label{extwobittensor}
Let qubit $A$ be prepared in state $\vert 0\rangle$, and qubit $B$ in state $\vert 1\rangle$. The state vector of the 2-qubit system $AB$ is $\vert 01\rangle=\vert 0\rangle\otimes\vert 1\rangle$, the tensor product of $\vert 0\rangle$ and $\vert 1\rangle$.
\end{example}

Let $Q_A=\left\{\vert q_0\rangle,\vert q_1\rangle,\ldots,\vert q_{N_A-1}\rangle\right\}$ and $R_B=\left\{\vert r_0\rangle,\vert r_1\rangle,\ldots,\vert r_{N_B-1}\rangle\right\}$ be orthonormal bases of $A$ and $B$, respectively. Then, due to the mathematical construction of tensor-product Hilbert spaces, $S=\left\{\vert q_i\rangle\otimes\vert r_j\rangle\,\vert\, 0\le i<N_A,0\le j<N_B\right\}$ is an orthonormal basis of $AB$.\footnote{Orthonormality is consistent with the physical fact that the product states in $S$ can be reliably distinguished from each other.} We can also write $\vert\alpha\rangle=\sum_{i=0}^{N_A-1}x_i\vert  q_i\rangle$ and $\vert\beta\rangle=\sum_{j=0}^{N_B-1}y_j\vert r_j\rangle$. As the tensor product $\otimes$ is linear in its arguments, $\vert\alpha\rangle\otimes\vert\beta\rangle=\sum_{i=0}^{N_A-1}\sum_{j=0}^{N_B-1}x_i y_j\left(\vert q_i\rangle\otimes\vert r_j\rangle\right)$. This shows that, in terms of basis $S$, the product state $\vert\alpha\rangle\otimes\vert\beta\rangle$ can be identified with the Kronecker product of the column vectors for $\vert\alpha\rangle$ and $\vert\beta\rangle$.

\begin{example}\label{extwoqubitbasis}
The 2-qubit computational basis is $S=\left\{\vert 00\rangle,\vert 01\rangle,\vert 10\rangle,\vert 11\rangle\right\}$, built from $Q_A=\left\{\vert 0\rangle,\vert 1\rangle\right\}$ and $R_B=\left\{\vert 0\rangle,\vert 1\rangle\right\}$. Thus, any 2-qubit state $\vert\psi\rangle$ can be written as a unique linear combination $\vert\psi\rangle=a_{00}\vert 00\rangle+a_{01}\vert 01\rangle+a_{10}\vert 10\rangle+a_{11}\vert 11\rangle$, with appropriate amplitudes. A 2-qubit system can be either in a classical state such as $\vert 01\rangle$, or in superposition of multiple classical states. In terms of $S$, the state $\vert 01\rangle$ is identified with the column vector $\begin{bmatrix}1 & 0\end{bmatrix}^T\otimes\begin{bmatrix}0 & 1\end{bmatrix}^T=\begin{bmatrix}0 & 1 & 0 & 0\end{bmatrix}^T$, where $\otimes$ is the Kronecker product.
\end{example}

\textbf{Entangled state.} States that cannot be written in the product form $\vert\alpha\rangle\otimes\vert\beta\rangle$ are called entangled states. Entanglement is a rich source of quantum weirdness.

\begin{example}\label{exentangled}
Perhaps the most famous 2-qubit entangled state is the \textit{EPR pair} $\vert\psi\rangle=\frac{1}{\sqrt{2}}\vert 00\rangle+\frac{1}{\sqrt{2}}\vert 11\rangle$. To show that the EPR pair is not a product state, let's assume for a moment that $\vert\psi\rangle=\vert\alpha\rangle\otimes\vert\beta\rangle$ holds, with some $\vert\alpha\rangle=x_0\vert 0\rangle+x_1\vert 1\rangle$ and $\vert\beta\rangle=y_0\vert 0\rangle+y_1\vert 1\rangle$. Then, we can write $\vert\psi\rangle=x_0y_0\vert 00\rangle+x_0y_1\vert 01\rangle+x_1y_0\vert 10\rangle+x_1y_1\vert 11\rangle$. But since the amplitudes are unique, $x_0y_0=x_1y_1=\frac{1}{\sqrt{2}}$ and $x_0y_1=x_1y_0=0$ must hold, which is impossible for any choice of $x_0,x_1,y_0,y_1$.
\end{example}

When a bipartite system $AB$ is in an entangled state, no state vector can be assigned to its subsystems $A$ and $B$ (otherwise, it would be a product state). That is, in the state-vector formalism it's possible that $AB$ as a whole has a quantum state, while its parts $A$ and $B$ do not!

\subsection{Two-qubit unitary transformations}\label{edu2qubitunitary}

Given the 2-qubit computational basis, any 2-qubit state $\vert\psi\rangle$ can be identified with a $4\times 1$ matrix (column vector) of amplitudes:
\begin{align}\label{eqtwoqubitcol}
\vert\psi\rangle &= \begin{bmatrix}
          a_{00} \\
          a_{01} \\
          a_{10} \\
          a_{11}
        \end{bmatrix}
\end{align}

Moreover, any 2-qubit quantum gate $U$ can be identified with a $4\times 4$ unitary matrix. A very important 2-qubit quantum gate is the CNOT gate, which is identified with:
\begin{align}\label{eqcnotmatrix}
CX &= \begin{bmatrix}
          1 & 0 & 0 & 0 \\
          0 & 1 & 0 & 0 \\
          0 & 0 & 0 & 1 \\
          0 & 0 & 1 & 0
        \end{bmatrix}
\end{align}

We may also apply a 1-qubit gate on a 2-qubit system. E.g. the effect of applying the Pauli-$X$ gate on the first qubit (while leaving the second alone) can be calculated using:
\begin{align}\label{eqcnotmatrix}
X\otimes I  &= \begin{bmatrix}
          0 & 1 \\
          1 & 0 
        \end{bmatrix} \otimes \begin{bmatrix}
          1 & 0 \\
          0 & 1
        \end{bmatrix} = \begin{bmatrix}
          0 & 0 & 1 & 0 \\
          0 & 0 & 0 & 1 \\
          1 & 0 & 0 & 0 \\
          0 & 1 & 0 & 0
        \end{bmatrix}
\end{align}

Here, $\otimes$ denotes the Kronecker product, and the 1-qubit gate $I$ represents "doing nothing" to the second qubit.\footnote{In general, $\left(U\otimes I\right)\left(\vert\alpha\rangle\otimes\vert\beta\rangle\right)=\left(U\vert\alpha\rangle\right)\otimes\left(I\vert\beta\rangle\right)=\left(U\vert\alpha\rangle\right)\otimes\vert\beta\rangle$. Thus, mathematically $U\otimes I$ behaves as expected for product states, bringing $\vert\alpha\rangle$ to $U\vert\alpha\rangle$ and leaving $\vert\beta\rangle$ alone. Now, assuming that "applying $U$ on the first subsystem" is physically indeed a unitary transformation of the whole, linearity implies that $U\otimes I$ must be a matrix representation of it, because any bipartite state can be written as a linear combination of product states.}

\begin{example}\label{expaulix1}
The state $\vert 00\rangle$ is identified with the column vector $\begin{bmatrix}1 & 0 & 0 & 0\end{bmatrix}^T$. Applying the Pauli-$X$ gate on the first qubit gives $\left(X\otimes I\right)\vert 00\rangle=\begin{bmatrix}0 & 0 & 1 & 0\end{bmatrix}^T=\vert 10\rangle$. As for the EPR pair which is identified with $\begin{bmatrix}\frac{1}{\sqrt{2}} & 0 & 0 & \frac{1}{\sqrt{2}}\end{bmatrix}^T$, we get $\left(X\otimes I\right)\left(\frac{1}{\sqrt{2}}\vert 00\rangle+\frac{1}{\sqrt{2}}\vert 11\rangle\right)=\begin{bmatrix}0 & \frac{1}{\sqrt{2}} & \frac{1}{\sqrt{2}} & 0\end{bmatrix}^T=\frac{1}{\sqrt{2}}\vert 10\rangle+\frac{1}{\sqrt{2}}\vert 01\rangle$. In the bra-ket notation, the results are intuitive, all we have to do is to negate the first bit in every bit string.
\end{example}

\begin{example}\label{exeprprepare}
Let's prepare the EPR pair. Starting with $\vert 00\rangle=\vert 0\rangle\otimes\vert 0\rangle$, formally the procedure is $CX\left(H\otimes I\right)\vert 00\rangle$. That is, first the Hadamard gate is applied on the first qubit: $\left(H\otimes I\right)\vert 00\rangle=\left(\frac{1}{\sqrt{2}}\vert 0\rangle+\frac{1}{\sqrt{2}}\vert 1\rangle\right)\otimes\vert 0\rangle=\frac{1}{\sqrt{2}}\vert 00\rangle+\frac{1}{\sqrt{2}}\vert 10\rangle$. From this, we get the EPR pair by applying the CNOT gate. To see why, either just do the matrix multiplication, or notice that $CX\vert 00\rangle=\vert 00\rangle$ and $CX\vert 10\rangle=\vert 11\rangle$. CNOT stands for "Controlled-NOT". When acting on a 2-qubit classical state, it negates the second (target) bit if and only if the first (control) bit is $1$.
\end{example}

\subsection{Partial measurement and the EPR problem}\label{edupartialepr}

Let $AB$ be a bipartite system and $Q_A=\left\{\vert q_0\rangle,\vert q_1\rangle,\ldots,\vert q_{N_A-1}\rangle\right\}$ an orthonormal basis of subsystem $A$. It can be proved\footnote{Using the $S=\left\{\vert q_i\rangle\otimes\vert r_j\rangle\right\}_{i,j}$ basis from Subsection \ref{edubipartite}.} that any state $\vert\psi\rangle$ of $AB$ can be written in the form \cite{mermin}:
\begin{multline}\label{eqborngen}
\vert\psi\rangle=a_0\vert q_0\rangle\otimes\vert\phi_0\rangle+a_1\vert q_1\rangle\otimes\vert\phi_1\rangle \\
+\ldots+a_{N_A-1}\vert q_{N_A-1}\rangle\otimes\vert\phi_{N_A-1}\rangle
\end{multline}

Here, each $\vert\phi_i\rangle$ is a unit vector, and for the $a_i$ values $\sum_{i=0}^{N_A-1}\vert a_i\vert^2=1$ holds.

If we now measure $A$ in basis $Q_A$,\footnote{Physically, imagine the "measurement-in-$Q_A$" protocol is executed on $A$, without touching $B$.} the outcome with respect to the whole system $AB$ is governed by the \textit{generalized Born rule} \cite{mermin}: with probability $\vert a_i\vert^2$ the state of $AB$ will collapse into $\vert q_i\rangle\otimes\vert\phi_i\rangle$, for every $0\le i<N_A$. (Each outcome has its distinct numerical result as well, but we ignore that in this paper.)

\begin{example}\label{exeprproblem}
Let's measure the first qubit of the EPR pair $\frac{1}{\sqrt{2}}\vert 00\rangle+\frac{1}{\sqrt{2}}\vert 11\rangle$, in the computational basis. The EPR pair is already in the form of Eq. \ref{eqborngen}, so the generalized Born rule can be directly applied. That is, with probability $\frac{1}{2}$ the 2-qubit state will collapse into $\vert 00\rangle$, and also with probability $\frac{1}{2}$ it will collapse into $\vert 11\rangle$. In the first case, $\vert 00\rangle=\vert 0\rangle\otimes\vert 0\rangle$ means both qubits are in state $\vert 0\rangle$ right after the measurement, while in the second case, $\vert 11\rangle=\vert 1\rangle\otimes\vert 1\rangle$ means both qubits are in state $\vert 1\rangle$.
\end{example}

Curiously, the previous example poses a serious problem. Imagine Alice possesses the first qubit and Bob the second (physically e.g. two atoms), and let they reside in two different galaxies, many light years from each other. If now Alice measures her qubit, then, depending on the outcome, the state of Bob's qubit collapses either into $\vert 0\rangle$ or $\vert 1\rangle$, basically instantaneously. So the impact of Alice's measurement seems to arrive at Bob faster than light.\footnote{Let's assume Alice and Bob are in the same inertial reference frame, to avoid clock-synchronization issues.}

To show why Alice cannot use the above effect to send a signal to Bob faster than light, next we'll present the density-matrix formalism.

\subsection{State as a density matrix}\label{edudensity}

In the following, a state vector $\vert\psi\rangle$ is always understood to be a column vector in terms of some pre-agreed orthonormal basis.

Let's consider a physical system with an $N$-dimensional associated state space.\footnote{As it was mentioned in Subsection \ref{edustatevec}, quantum theory postulates a state space always exists, and it's an $N$-dimensional complex Hilbert space.} Any positive semi-definite $N\times N$ complex matrix $\rho$ with $\mathrm{tr}\left(\rho\right)=1$ (i.e. trace one) is called a \textit{density matrix} of the system. It turns out that besides state vectors, density matrices can also be used to describe the state of quantum systems, even in situations where state vectors cannot.

In the \textit{density-matrix formalism}, the idea is that in certain situations, it is a density matrix $\rho$ that lets us calculate the outcome probabilities for any conceivable measurement on the system, and because of that, $\rho$ can be thought of as the quantum state \cite{schumacher,hardy}. In particular, if we measure the system in an (orthonormal) basis $B=\left\{\vert b_0\rangle,\vert b_1\rangle,\ldots,\vert b_{N-1}\rangle\right\}$, the probability of collapsing into $\vert b_i\rangle$ is given by $\langle b_i\vert\rho\vert b_i\rangle$, for each $0\le i<N$.

Density matrices arise in the following situations:

\textbf{Pure state.} We prepare the system in state $\vert\psi\rangle$. The density matrix for this situation is $\rho=\vert\psi\rangle\langle\psi\vert$,\footnote{This is an outer product, multiplying the $N\times 1$ matrix $\vert\psi\rangle$ with the $1\times N$ matrix $\langle\psi\vert$.} which is called a pure state, representing the very same physical state as $\vert\psi\rangle$ does. If we measure in basis $B$, the probability of collapsing into $\vert b_i\rangle$ is $P_i=\langle b_i\vert\psi\rangle\langle\psi\vert b_i\rangle=\langle b_i\vert\rho\vert b_i\rangle$, see Subsection \ref{edumanipulate}. 

\textbf{Mixture of pure states.} We are given a system that was prepared with probability $p_1$ in state $\vert\psi_1\rangle$, with $p_2$ in $\vert\psi_2\rangle$, $\ldots$ , with $p_m$ in $\vert\psi_m\rangle$, where $m\ge 1$ and $p_1+p_2+\ldots+p_m=1$.

The density matrix describing this situation from our perspective is $\rho=\sum_{k=1}^m p_k\vert\psi_k\rangle\langle\psi_k\vert$. It can be checked via the calculation below that the probability of collapsing into $\vert b_i\rangle$ is indeed $\langle b_i\vert\rho\vert b_i\rangle$, if we measure in basis $B$:
\begin{multline}\label{eqmixedprob}
P_i=\sum_{k=1}^m p_k\langle b_i\vert\psi_k\rangle\langle\psi_k\vert b_i\rangle \\
=\langle b_i\vert\underbrace{\left(\sum_{k=1}^m p_k\vert\psi_k\rangle\langle\psi_k\vert\right)}_{\rho}\vert b_i\rangle \\
=\langle b_i\vert\rho\vert b_i\rangle
\end{multline}

\begin{example}\label{exprepare01}
We are given a qubit which is either in state $\vert 0\rangle$ or $\vert 1\rangle$, with probability $\frac{1}{2}$ each. This means that $m=2$, $p_1=p_2=\frac{1}{2}$, $\vert\psi_1\rangle=\vert 0\rangle$, $\vert\psi_2\rangle=\vert 1\rangle$. In the computational basis, the resulting density matrix is $\rho=\frac{1}{2}\vert 0\rangle\langle 0\vert+\frac{1}{2}\vert 1\rangle\langle 1\vert=\frac{1}{2}\begin{bmatrix}1\\0\end{bmatrix}\begin{bmatrix}1 & 0\end{bmatrix}+\frac{1}{2}\begin{bmatrix}0\\1\end{bmatrix}\begin{bmatrix}0 & 1\end{bmatrix}=\frac{1}{2}\begin{bmatrix}1 & 0\\0 & 1\end{bmatrix}=\frac{I}{2}$, the so-called \textit{1-qubit maximally mixed state}. Now, if we measure in basis $B=\left\{\vert b_0\rangle,\vert b_1\rangle\right\}$, the collapse probabilities are $P_0=\langle b_0\vert\rho\vert b_0\rangle=\frac{1}{2}$ and $P_1=\langle b_1\vert\rho\vert b_1\rangle=\frac{1}{2}$. That is, the probability distribution is the same for any basis we may measure in.
\end{example}

Mathematically, a pure state $\rho=\vert\psi\rangle\langle\psi\vert$ cannot be written as a mixture $\rho=\sum_{k=1}^m p_k\vert\psi_k\rangle\langle\psi_k\vert$, unless $\vert\psi_k\rangle=c_k\vert\psi\rangle$ for all $p_k>0$, where $c_k$ is a complex number with $\lvert c_k\rvert=1$. In the density-matrix formalism, a non-pure state is called a \textit{mixed state}.

Using its spectral decomposition, \textit{any} density matrix $\rho$ can be prepared as a mixture of orthogonal pure states.

\textbf{Reduced state.} We are given subsystem $A$ of a bipartite system $AB$ whose state $\rho_{AB}$ is either a pure state or a mixture of pure states.

It can be proved\footnote{Using the generalized Born rule from Subsection \ref{edupartialepr}, which determines the outcome probabilities for any measurement on a subsystem.} that the density matrix for this situation is $\rho_A=\mathrm{tr}_B\left(\rho_{AB}\right)$, i.e. the partial trace of $\rho_{AB}$ over $B$. In other words, if we know $\rho_{AB}$, we can calculate $\rho_A$ from it.

\begin{example}\label{exprepare01viaepr}
One method to work out the partial trace over $B$ is to assume that $B$ has already been measured in some basis $R_B$, but we don't know what the outcome was. Then, we calculate the density matrix for $A$, from our perspective. It doesn't matter which basis $R_B$ is used, the end result will always be the same, and that's the $\rho_A$ we are looking for. Let's try it with the EPR pair $\frac{1}{\sqrt{2}}\vert 00\rangle+\frac{1}{\sqrt{2}}\vert 11\rangle$. If someone (hypothetically) measured the second qubit $B$ in the computational basis,\footnote{An equation analogous to Eq. \ref{eqborngen} can be derived for subsystem $B$ with basis $R_B$.} then from our perspective, the first qubit $A$ would become $\vert 0\rangle$ with probability $\frac{1}{2}$ and $\vert 1\rangle$ with probability $\frac{1}{2}$ as well, which is a mixture of the pure states $\vert 0\rangle$ and $\vert 1\rangle$. Therefore, the density matrix of the first qubit of the EPR pair is $\rho_A=\frac{1}{2}\vert 0\rangle\langle 0\vert+\frac{1}{2}\vert 1\rangle\langle 1\vert$, which is $\frac{I}{2}$ in the computational basis.
\end{example}

In principle, \textit{any} density matrix $\rho_A$ can be prepared this way, i.e. it's always possible to set up an extended system $AB$ such that $\rho_A=\mathrm{tr}_B\left(\rho_{AB}\right)$ holds. Here, $B$ is called the \textit{ancilla system}, or ancilla for short, as its only purpose is to help prepare $\rho_A$.

\subsection{Classical vs. quantum mixture}\label{educvqmixture}

In Example \ref{exprepare01}, one might argue that the state of the qubit isn't $\rho=\frac{1}{2}\vert 0\rangle\langle 0\vert+\frac{1}{2}\vert 1\rangle\langle 1\vert$, but rather one of $\vert 0\rangle$ or $\vert 1\rangle$, we just don't know which. It seems counter-intuitive that in the density-matrix formalism our ignorance about the exact state vector is incorporated in the quantum state. To justify why it is done like that, the following two examples highlight a crucial difference between classical and quantum mixtures.

\begin{example}\label{exrandomcoin}
Bob is given a (classical) coin in a closed box, knowing nothing about how the coin was "prepared", i.e. whether it is heads or tails. If Bob isn't allowed to open up the box, his \textit{logical} situation includes his ignorance. However, physically Bob can always open up the box and observe the exact state of the coin. That is to say, in the classical world Bob cannot be in a \textit{physical} situation in his local lab such that he is fundamentally uncertain about the state of a system he has access to.
\end{example}

\begin{example}\label{exquantumcoin}
Bob is given a qubit, knowing nothing about how it was prepared. Let's assume the qubit was prepared by a random process, either in state $\vert 0\rangle$ or $\vert 1\rangle$, each with probability $\frac{1}{2}$. However, if Bob measures the qubit in the computational basis and gets $0$, he cannot conclude that the qubit was originally in the $\vert 0\rangle$ state. From Bob's perspective, it's also possible that the qubit was e.g. in state $0.6\vert 0\rangle+0.8\vert 1\rangle$, or it could have been the first qubit of an EPR pair, which doesn't even have its own state vector! There is no way he can resolve this uncertainty about the state vector (or lack of it) on his own, by applying operations allowed in quantum mechanics.
\end{example}

In the previous example, if Bob was given a large sample of instances, it would enable him to find out (estimate) the density matrix of yet unseen instances, using a procedure called quantum-state tomography \cite{nielsen}.

In general, the density matrix represents a quantum system in itself, encapsulating all that an experimenter can (statistically) find out by accessing a large sample of instances.

\subsection{Distinguishability and the no-signaling principle}\label{edutellapart}

Let's apply a unitary transformation, identified with the unitary matrix $U$, on a system which has density matrix $\rho$.

It can be shown that the new density matrix will be $\rho'=U\rho U^{\dagger}$, where $U^{\dagger}$ is the conjugate transpose of $U$. That is, $\rho'$ can be used after the transformation to calculate outcome probabilities for measurements.

\begin{example}\label{exunitaryon01}
The qubit in Example \ref{exprepare01} is either $\vert 0\rangle$ or $\vert 1\rangle$, with probability $\frac{1}{2}$ each. The density matrix is therefore $\rho=\frac{1}{2}\vert 0\rangle\langle 0\vert+\frac{1}{2}\vert 1\rangle\langle 1\vert$. If we now apply a Pauli-$X$ gate, then the qubit will be either $X\vert 0\rangle$ or $X\vert 1\rangle$, with probability $\frac{1}{2}$ each, where $X$ denotes the unitary matrix of the Pauli-$X$ gate, in the computational basis. So the new density matrix is $\rho'=\frac{1}{2}X\vert 0\rangle\langle 0\vert X^{\dagger}+\frac{1}{2}X\vert 1\rangle\langle 1\vert X^{\dagger}=\frac{1}{2}\vert 1\rangle\langle 1\vert+\frac{1}{2}\vert 0\rangle\langle 0\vert$.\footnote{For a mixture of pure states, the new state vector is $U\vert\psi_k\rangle$ with probability $p_k$, so the new density matrix is $\rho'=\sum_{k=1}^m p_k U\vert\psi_k\rangle\langle\psi_k\vert U^{\dagger}=U\left(\sum_{k=1}^m p_k \vert\psi_k\rangle\langle\psi_k\vert\right) U^{\dagger}=U\rho U^{\dagger}$, where $\langle\psi_k\vert U^{\dagger}$ is the conjugate transpose of $U\vert\psi_k\rangle$. The formula $\rho'=U\rho U^{\dagger}$ can be shown to hold for reduced states as well.} Since $\rho'=\rho$, the density matrix hasn't changed, although we did apply a non-trivial transformation.
\end{example}

Two quantum systems with the same density matrix $\rho_1=\rho_2$ behave exactly the same way (w.r.t. quantum operations) for an experimenter who knows only the density matrix. On the other hand, if $\rho_1\ne\rho_2$, the systems can always be told apart, in the sense that applying well-chosen operations would always result in an increased confidence in differentiating.

\begin{example}\label{extellapart}
Let $\rho_1=\vert 0\rangle\langle 0\vert$ and $\rho_2=0.99\vert 0\rangle\langle 0\vert+0.01\vert 1\rangle\langle 1\vert$, and let's assume we are given equally likely one instance of either $\rho_1$ or $\rho_2$, but we don't know which. Now, if we measure in the computational basis and get $1$, we can be sure we've got an instance of $\rho_2$. On the other hand, if the result is $0$, our guess would be $\rho_1$ and we'd have more than $50\%$ chance to be correct (an increased confidence compared to the initial $50$-$50\%$). If we are given a large number of instances instead of just one, we'd succeed with high confidence, as even a single $1$ among the measurement results would indicate with certainty that we were given instances of $\rho_2$; otherwise, if all results are $0$, we can be almost sure we were given instances of $\rho_1$.
\end{example}

Now it's easy to see why Alice cannot send a signal to Bob by measuring her qubit of an EPR pair (see Example \ref{exeprproblem}). It is because, as Example \ref{exprepare01viaepr} shows, the quantum state of Bob's qubit of the EPR pair is the 1-qubit maximally mixed state, and from Bob's perspective it will remain the same even after Alice has measured her qubit.

If Alice and Bob decide to use a large number of EPR pairs instead of just one, Bob still wouldn't be able to tell whether Alice has measured all her qubits or none of them, because examining Bob's qubits would yield statistics that come from the same underlying probability distribution in both situations. So Alice cannot send in this way a signal to Bob instantaneously, i.e. faster than light.

The \textit{no-signaling principle} is a no-go theorem of quantum information, saying that no bipartite quantum state can be used to send information instantaneously (other than just random $0$s and $1$s).

\subsection{Bipartite systems revisited}\label{eduentwdensities}

Using density matrices to represent quantum states, the state $\rho_{AB}$ of a bipartite system $AB$ can be categorized as follows.

\textbf{Product state.} If $A$ and $B$ are prepared independently, in state $\rho_A$ and $\rho_B$, respectively, it can be proved that the overall density matrix is the Kronecker product $\rho_{AB}=\rho_A\otimes\rho_B$. As for the partial trace, $\rho_A=\mathrm{tr}_B\left(\rho_A\otimes\rho_B\right)$, as one would intuitively expect.

\textbf{Separable state.} This is a generalization of the product state. Formally, $\rho_{AB}$ is a separable state when:
\begin{equation}\label{eqseparable}
\rho_{AB}=\sum_{k=1}^m p_k\left(\rho_{A,k}\otimes\rho_{B,k}\right)
\end{equation}

Here, $m\ge 1$, each $p_k\ge 0$ is a real number, and $p_1+p_2+\ldots+p_m=1$. One way to prepare $\rho_{AB}$ is to prepare $AB$ with probability $p_1$ in the product state $\rho_{A,1}\otimes\rho_{B,1}$, with $p_2$ in $\rho_{A,2}\otimes\rho_{B,2}$, $\ldots$ , with $p_m$ in $\rho_{A,m}\otimes\rho_{B,m}$, that is, a mixture of product states.

Again, the partial trace is intuitive: $\rho_A=\mathrm{tr}_B\left(\rho_{AB}\right)=\sum_{k=1}^m p_k\rho_{A,k}$.

\begin{example}\label{exseparable}
The 2-qubit state $\rho_{AB}=\frac{1}{2}\vert 00\rangle\langle 00\vert+\frac{1}{2}\vert 11\rangle\langle 11\vert$ is a separable state, because using matrix algebra it can be written as $\rho_{AB}=\frac{1}{2}\vert 0\rangle\langle 0\vert\otimes\vert 0\rangle\langle 0\vert+\frac{1}{2}\vert 1\rangle\langle 1\vert\otimes\vert 1\rangle\langle 1\vert$. From this, we can calculate that $\rho_{A,1}=\rho_{B,1}=\vert 0\rangle\langle 0\vert$, $\rho_{A,2}=\rho_{B,2}=\vert 1\rangle\langle 1\vert$, $\rho_A=\mathrm{tr}_B\left(\rho_{AB}\right)=\frac{1}{2}\vert 0\rangle\langle 0\vert+\frac{1}{2}\vert 1\rangle\langle 1\vert$, $\rho_B=\mathrm{tr}_A\left(\rho_{AB}\right)=\frac{1}{2}\vert 0\rangle\langle 0\vert+\frac{1}{2}\vert 1\rangle\langle 1\vert$. However, $\rho_{AB}\ne\rho_A\otimes\rho_B$ indicates that $\rho_{AB}$ isn't a product state, so it cannot be prepared by preparing $A$ and $B$ independently. E.g. if we prepare $\rho_{AB}$ as a mixture of product states, coordination is necessary to make sure $A$ and $B$ are prepared in the matching states $\rho_{A,k}$ and $\rho_{B,k}$, respectively, with a randomly picked $k\in\left\{1,2\right\}$.
\end{example}

\begin{example}\label{exseparable2}
The 2-qubit state $\rho_{AB}=\frac{1}{2}\vert 00\rangle\langle 00\vert+\frac{1}{2}\vert 11\rangle\langle 11\vert$ can also be prepared as the 2-qubit subsystem $AB$ of a 3-qubit system $ABC$ whose state vector is $\vert\psi_{ABC}\rangle=\frac{1}{\sqrt{2}}\vert 000\rangle+\frac{1}{\sqrt{2}}\vert 111\rangle$. Along the lines of Example \ref{exprepare01viaepr}, notice that if someone (hypothetically) measured the third qubit $C$, the generalized Born rule implies that the state of the first two qubits would collapse into either $\vert 00\rangle$ or $\vert 11\rangle$, with probability $\frac{1}{2}$ each. This means that the density matrix of the first two qubits in $ABC$ is the $\rho_{AB}$ given above.
\end{example}

\textbf{Entangled state.} If $\rho_{AB}$ cannot be written in the separable-state form of Equation \ref{eqseparable}, it is called an entangled state. In other words, a state is entangled if it cannot be prepared as a mixture of product states.

\begin{example}\label{exetangleddensity}
Unsurprisingly, the density matrix of the EPR pair is entangled. The EPR pair is a 2-qubit system $AB$, in a pure state with state vector $\vert\psi\rangle=\frac{1}{\sqrt{2}}\vert 00\rangle+\frac{1}{\sqrt{2}}\vert 11\rangle$. Accordingly, its density matrix is $\rho_{AB}=\vert\psi\rangle\langle\psi\vert=\frac{1}{2}\left(\vert 00\rangle+\vert 11\rangle\right)\left(\langle 00\vert+\langle 11\vert\right)$. We saw in Subsection \ref{edudensity} that basically this is the only way to write $\rho_{AB}$ as a mixture of pure states. Now, if $\rho_{AB}$ could be written as a mixture of product states, it would imply a mixture of \textit{pure} product states, which isn't possible, as in general $\vert\phi_A\rangle\otimes\vert\phi_B\rangle\ne c\vert\psi\rangle$.
\end{example}

\section{Theory - Part 1}\label{sec2}

Throughout the rest of the paper, we work with a bipartite quantum system $AB$, where subsystem $A$ belongs to Alice, and subsystem $B$ to Bob. The overall quantum state of $AB$ is represented by the density matrix $\rho_{AB}$, while those of $A$ and $B$ are given by the partial trace formulas $\rho_A=\mathrm{tr}_B( \rho_{AB})$ and $\rho_B=\mathrm{tr}_A( \rho_{AB})$, respectively. State vectors are understood to be column vectors, quantum states of 1- or 2-qubit systems are meant in terms of the computational basis, and the $\otimes$ symbol denotes the Kronecker product.

Let's start with the technical definition:

\begin{definition}[Quantum-house effect]\label{defqhe}
The quantum-house effect is the phenomenon when an operation on subsystem $A$ changes the overall state of system $AB$, but not that of $A$.
\end{definition}

\begin{figure}[t]
\centering
\includegraphics[width=1.0\columnwidth,trim={1.55cm 0 1.55cm 0},clip]{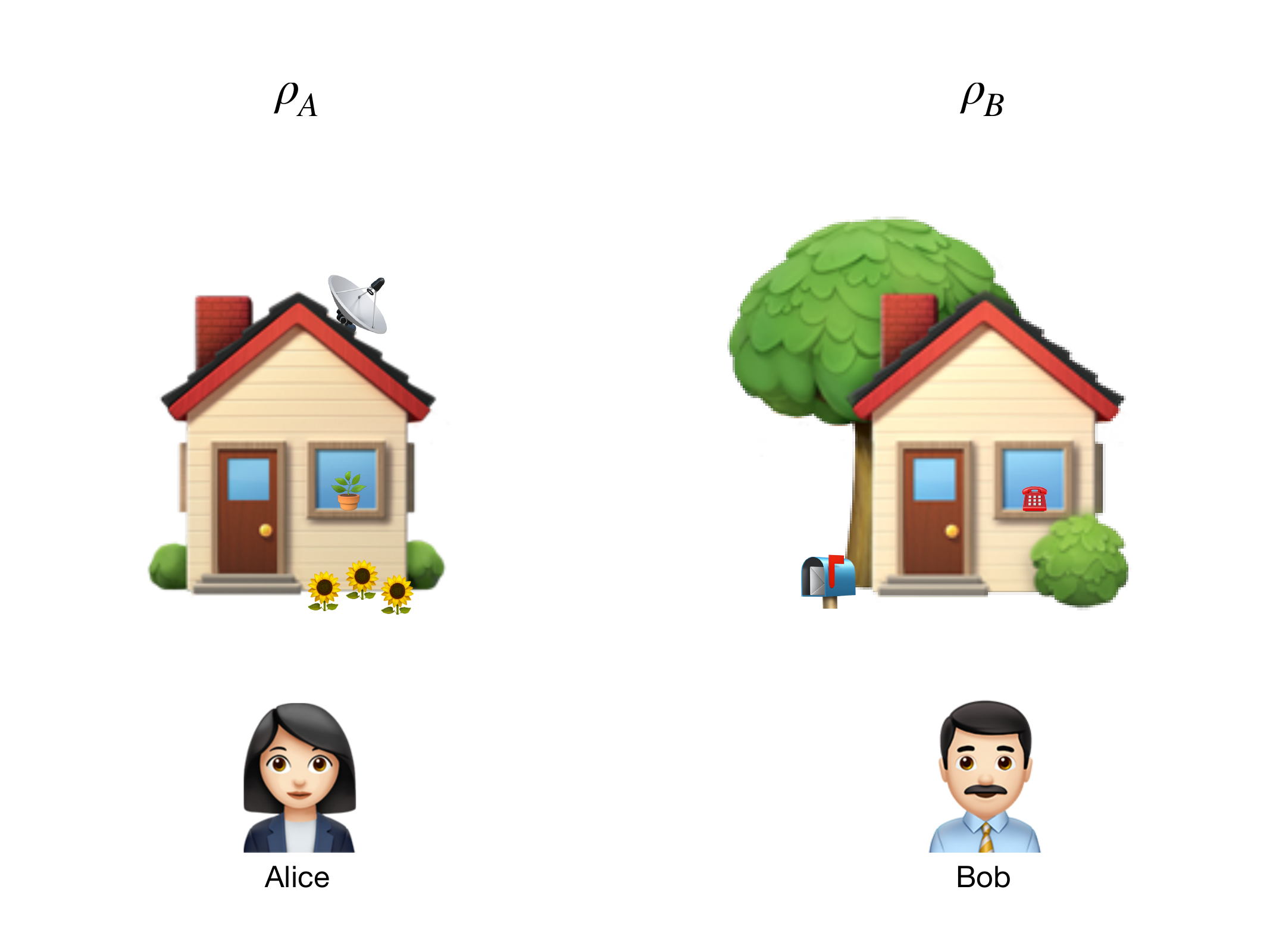}
\caption{Analogy for the quantum-house effect. Charlie enters Alice's house and either makes a change or does nothing. However, Alice cannot figure out, by examining solely her own house, whether or not Charlie has made any change. But if she joins forces with Bob, the two of them \textbf{together} may figure it out. This is because the change is such that it affects only the joint state $\rho_{AB}$ of the two houses, but not the state $\rho_{A}$ of Alice's individual house.}\label{figaqh}
\end{figure}

That is, the state of $AB$ changes to some $\rho'_{AB}\ne\rho_{AB}$, while the state of $A$ remains $\rho_A$. The "operation on $A$" can be anything, as long as it's performed inside Alice's lab: measurements, unitaries, or combinations thereof, with or without ancilla system.\footnote{When only unitaries are allowed without ancilla, such an operation not affecting the state of $A$ is called a locally non-effective unitary operation, see \cite{fu} and \cite{datta} for details.} The only restriction is that during the operation we don't have access to Bob's lab where subsystem $B$ resides.

The quantum-house effect is a non-classical phenomenon, because in the classical world any such operation in Alice's lab which doesn't change the state of $A$ cannot change the state of $AB$ either. (By "state" in the classical world we mean a \textit{complete} and \textit{objective} description of a physical system in itself that can be in principle established by an experimenter who is given access to the system in her local lab.) Furthermore, it is also a non-local effect in the sense that the impact of the operation spans two locations, i.e. labs, although according to our classical intuition it should be confined just to Alice's lab.

\begin{example}\label{exqhe}
Imagine that Charlie measures, in the computational basis, Alice's qubit of an EPR pair $\rho_{AB}=\frac{1}{2}(\vert 00\rangle+\vert 11\rangle)(\langle 00\vert+\langle 11\vert)$. He tells Alice and Bob that he performed the measurement, but doesn't tell anyone what the outcome was. Thus, for Alice and Bob, $AB$ is either in the state $\vert 00\rangle$ or $\vert 11\rangle$, with probability $\frac{1}{2}$ each, i.e. we can write $\rho'_{AB}=\frac{1}{2}\vert 00\rangle\langle 00\vert+\frac{1}{2}\vert 11\rangle\langle 11\vert$, which is different from $\rho_{AB}$. However, the state of Alice's individual qubit stays the same: $\rho'_A=\rho_A=\frac{1}{2}\vert 0\rangle\langle 0\vert+\frac{1}{2}\vert 1\rangle\langle 1\vert$, which is $\frac{I}{2}$, the 1-qubit maximally mixed state. So, in case of any doubt, Alice has no chance to figure out by herself whether Charlie has \textit{really} made the measurement. But \textit{together} with Bob, they may figure it out!
\end{example}

For a more intuitive understanding, an analogy is shown in Fig.~\ref{figaqh}. In this analogy, subsystems $A$ and $B$ are houses of Alice and Bob, respectively. Then, a change made secretly by Charlie on Alice's house may only be detected by Alice and Bob together, but not by Alice alone examining her own house.

The quantum-house effect can also be achieved with non-entangled states, as it can be seen in the next example. Thus, it extends the notion of quantum nonlocality to a wider range of bipartite quantum states than that offered by entanglement, and shows that already separable quantum states can behave in a counter-intuitive way in this regard.

\begin{example}\label{exqhezd}
Let $\rho_{AB}=\frac{1}{2}\vert 00\rangle\langle 00\vert+\frac{1}{2}\vert 11\rangle\langle 11\vert$. This is a separable, i.e. non-entangled state, since it can be written as $\rho_{AB}=\frac{1}{2}\vert 0\rangle\langle 0\vert\otimes\vert 0\rangle\langle 0\vert+\frac{1}{2}\vert 1\rangle\langle 1\vert\otimes\vert 1\rangle\langle 1\vert$. Now, if we apply a Pauli-$X$ gate on the first qubit, the overall 2-qubit state will change to $\rho'_{AB}=\frac{1}{2}\vert 10\rangle\langle 10\vert+\frac{1}{2}\vert 01\rangle\langle 01\vert$, which is different from $\rho_{AB}$. On the other hand, the state of the first qubit remains $\rho'_A=\rho_A=\frac{1}{2}\vert 0\rangle\langle 0\vert+\frac{1}{2}\vert 1\rangle\langle 1\vert$.
\end{example}

\section{Demonstration on SpinQ Gemini}\label{sec3}

In this section, we'll showcase the quantum-house effect on the SpinQ Gemini 2-qubit NMR desktop quantum computer \cite{hou}.

\begin{figure}[t]
\centering
\includegraphics[width=0.95\columnwidth]{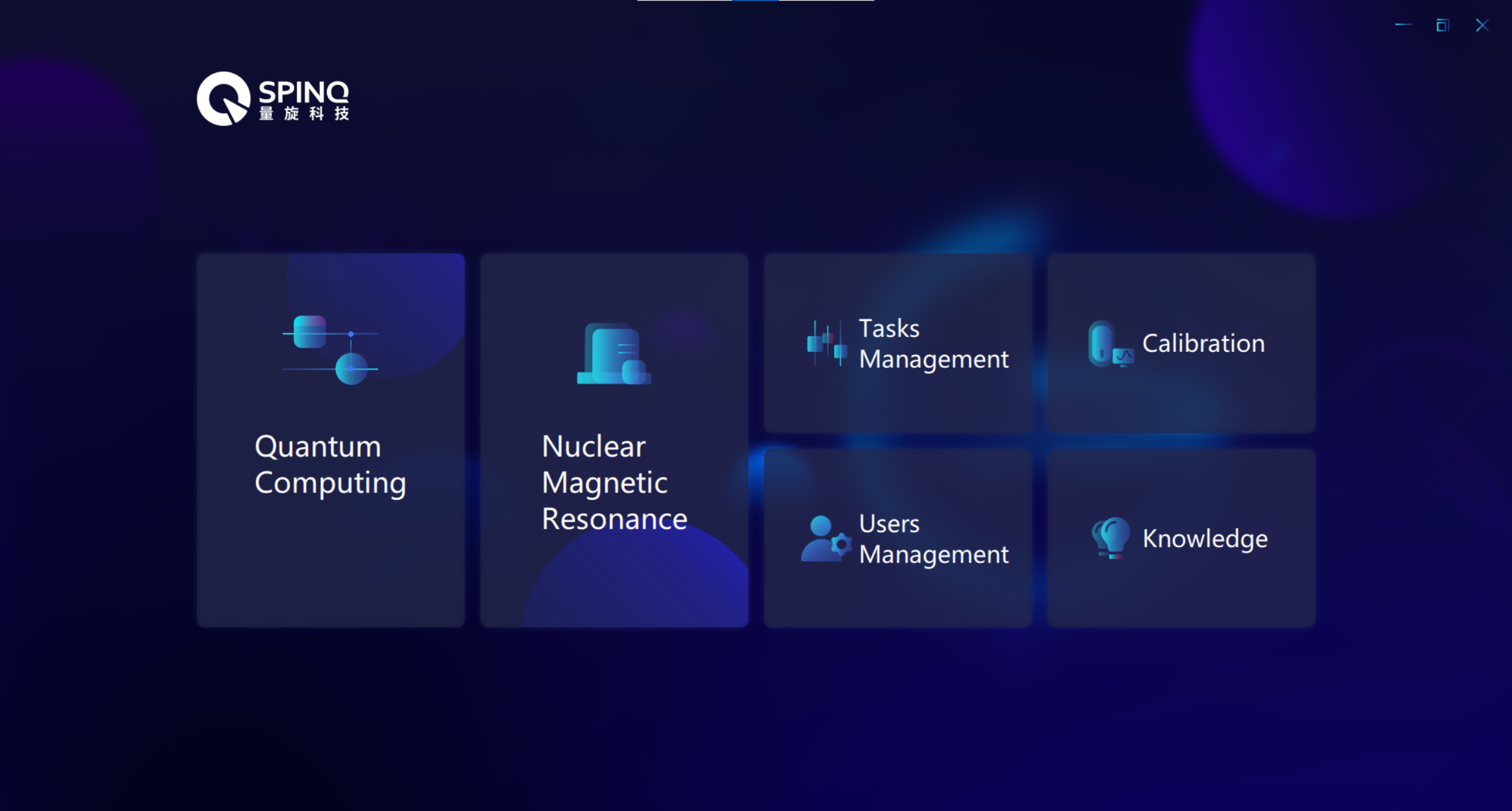}
\medskip
\caption{The homepage of SpinQuasar, the user-interface software for SpinQ Gemini, installed on a personal computer (PC).}\label{figquasar}
\end{figure}

The SpinQ Gemini device comes with the user-interface software SpinQuasar (see Fig.~\ref{figquasar}), together forming an integrated hardware-software platform for quantum computing education and research. For further technical details, including how the qubits are physically realized, the reader is referred to \cite{hou}.

We are going to demonstrate the following example on SpinQ Gemini, with the help of SpinQuasar:

\begin{example}\label{exgemini}
Charlie applies a Pauli-$X$ gate on Alice's qubit of an EPR pair $\rho_{AB}=\frac{1}{2}(\vert 00\rangle+\vert 11\rangle)(\langle 00\vert+\langle 11\vert)$, but doesn't tell anyone that he did so. Then, the overall 2-qubit state for Alice and Bob changes to $\rho'_{AB}=\frac{1}{2}(\vert 10\rangle+\vert 01\rangle)(\langle 10\vert+\langle 01\vert)$. However, the state of Alice's individual qubit stays the same: $\rho'_A=\rho_A=\frac{1}{2}\vert 0\rangle\langle 0\vert+\frac{1}{2}\vert 1\rangle\langle 1\vert$. So Alice has no chance to figure out by herself that Charlie did something. But \textit{together} with Bob, they may figure it out!
\end{example}

This example is similar to Example~\ref{exqhe}, but here Charlie (secretly) applies a unitary on Alice's qubit, instead of measuring it.

\begin{figure*}[p]
\centering
\includegraphics[width=0.98\linewidth]{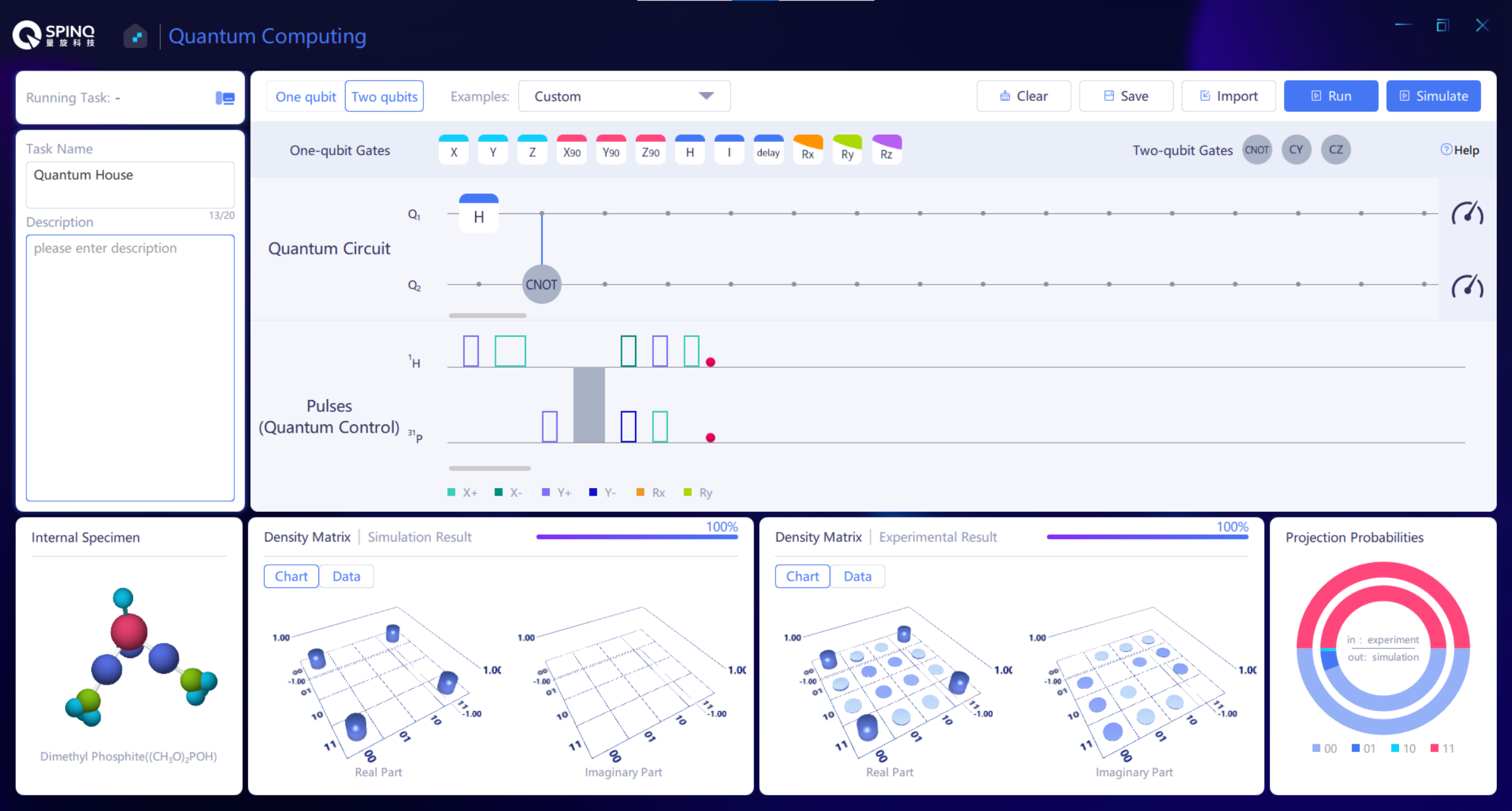}\\
\bigskip
\includegraphics[width=0.98\linewidth]{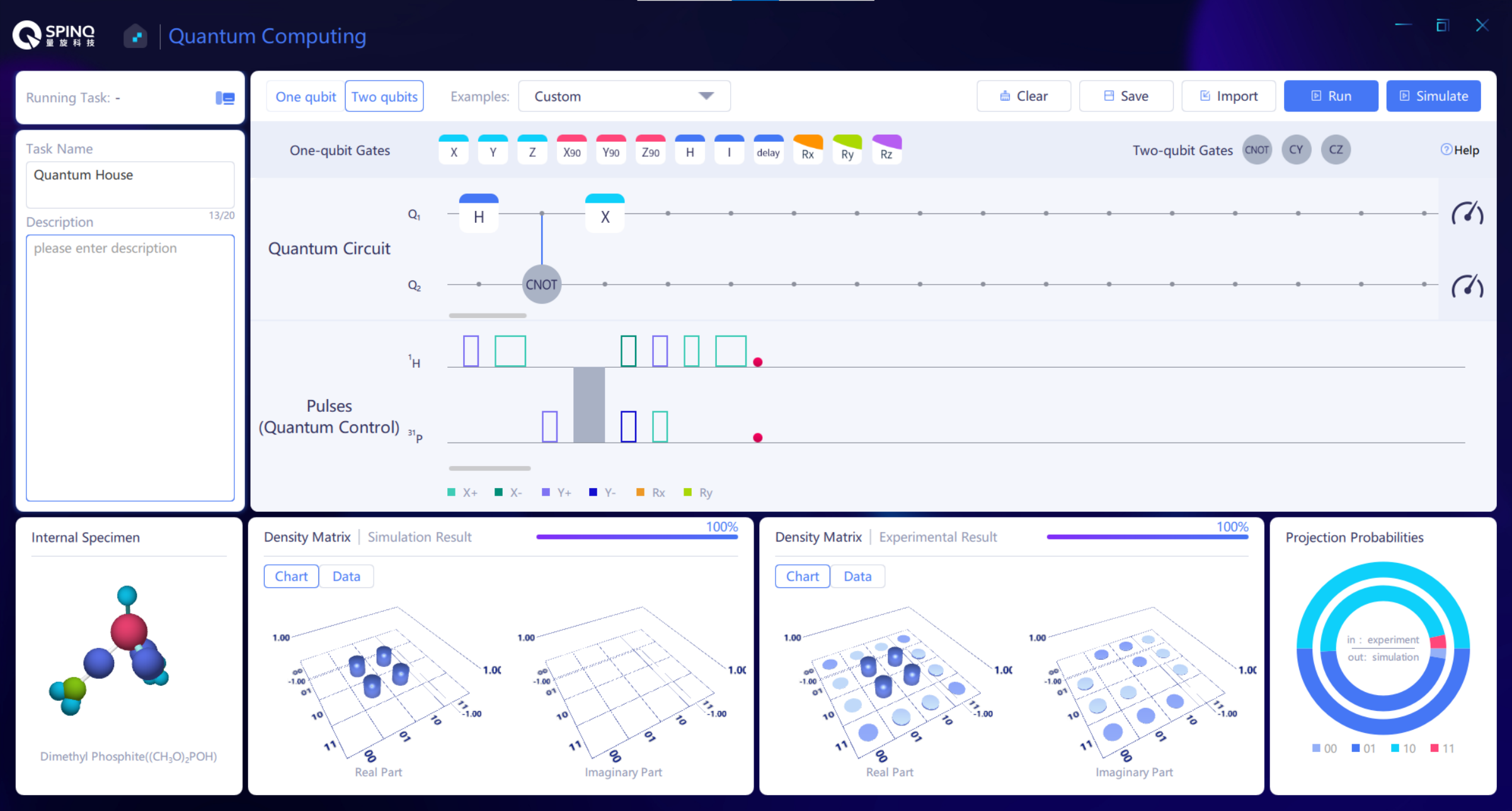}\\
\medskip
\caption{Overall impact of applying a Pauli-$X$ gate on the first qubit of the EPR pair $\rho_{AB}=\frac{1}{2}(\vert 00\rangle+\vert 11\rangle)(\langle 00\vert+\langle 11\vert)$. \textbf{Top screenshot:} preparation of the EPR pair $\rho_{AB}$ in SpinQ Gemini. In the "Quantum Circuit" section, there is a Hadamard gate ($H$) on the first qubit, followed by a CNOT gate; this circuit is executed to produce $\rho_{AB}$. The "Density Matrix $\vert$ Simulation Result" section shows the ideal $\rho_{AB}$ in chart format (see Fig.~\ref{figepr2qdata} for numerical format), where four matrix elements (in the four corners) have the value $0.5$, the rest are all zeros. Next to it, the "Density Matrix $\vert$ Experimental Result" section shows the noisy EPR-pair density matrix that is actually produced by the SpinQ Gemini hardware. In spite of the noise, the matrix elements in the four corners are still close to $0.5$, and the rest are all close to zero as well. \textbf{Bottom screenshot:} the 2-qubit state after applying an additional Pauli-$X$ gate on the first qubit of the EPR pair. In the density-matrix charts we can see a clear difference: the $0.5$ values have moved from the corners to the middle. The new ideal state is $\rho'_{AB}=\frac{1}{2}(\vert 10\rangle+\vert 01\rangle)(\langle 10\vert+\langle 01\vert)$, which is different from $\rho_{AB}$.}\label{figepr2q}
\end{figure*}

The SpinQuasar screenshots in Fig.~\ref{figepr2q} show how we implemented the EPR pair on SpinQ Gemini, as well as the EPR pair followed by a Pauli-$X$ gate on the first qubit. In each case, SpinQuasar displays not only the ideal, i.e. noiseless, 2-qubit density matrix, but also the noisy density matrix which was actually produced by the hardware.\footnote{Due to the peculiarities of liquid-state NMR technology, whenever we command SpinQ Gemini to produce a pure $n$-qubit state $\rho=\vert\psi\rangle\langle\psi\vert$, such as the EPR pair, the hardware will instead produce a so-called pseudo-pure state $\sigma=(1-\eta)\frac{I}{2^n}+\eta\vert\psi\rangle\langle\psi\vert$, where $\eta\sim 10^{-5}$ for $n=1,2$. This happens under the hood, and as $\rho$ and $\sigma$ are equivalent in the sense that we can unambiguously calculate one from the other, SpinQuasar only shows us $\rho$ (both ideal and noisy), but not $\sigma$.} We can clearly see that applying a Pauli-$X$ gate on the first qubit changes the overall 2-qubit state.

\begin{figure*}[p]
\centering
\includegraphics[width=0.98\linewidth]{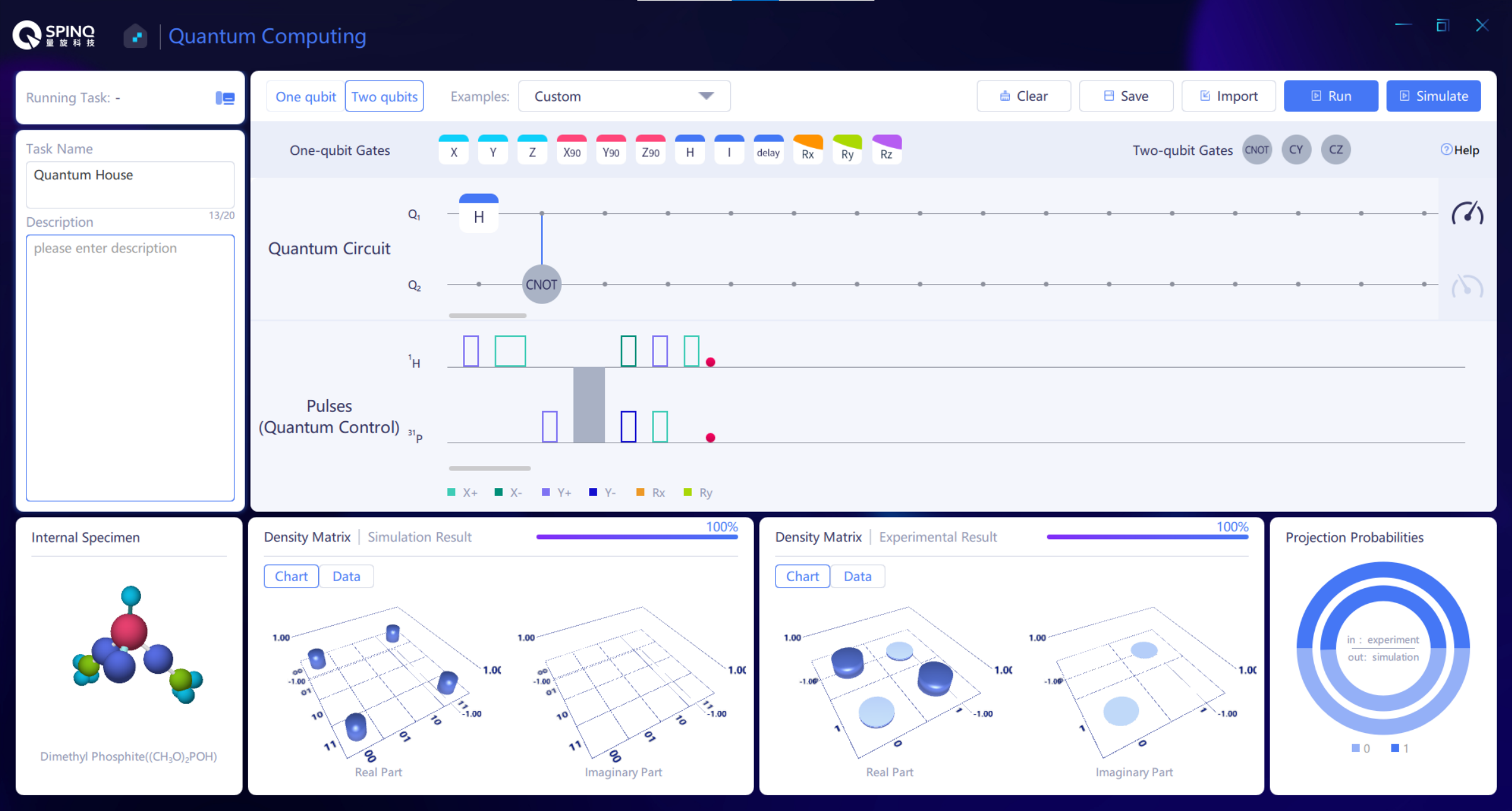}\\
\bigskip
\includegraphics[width=0.98\linewidth]{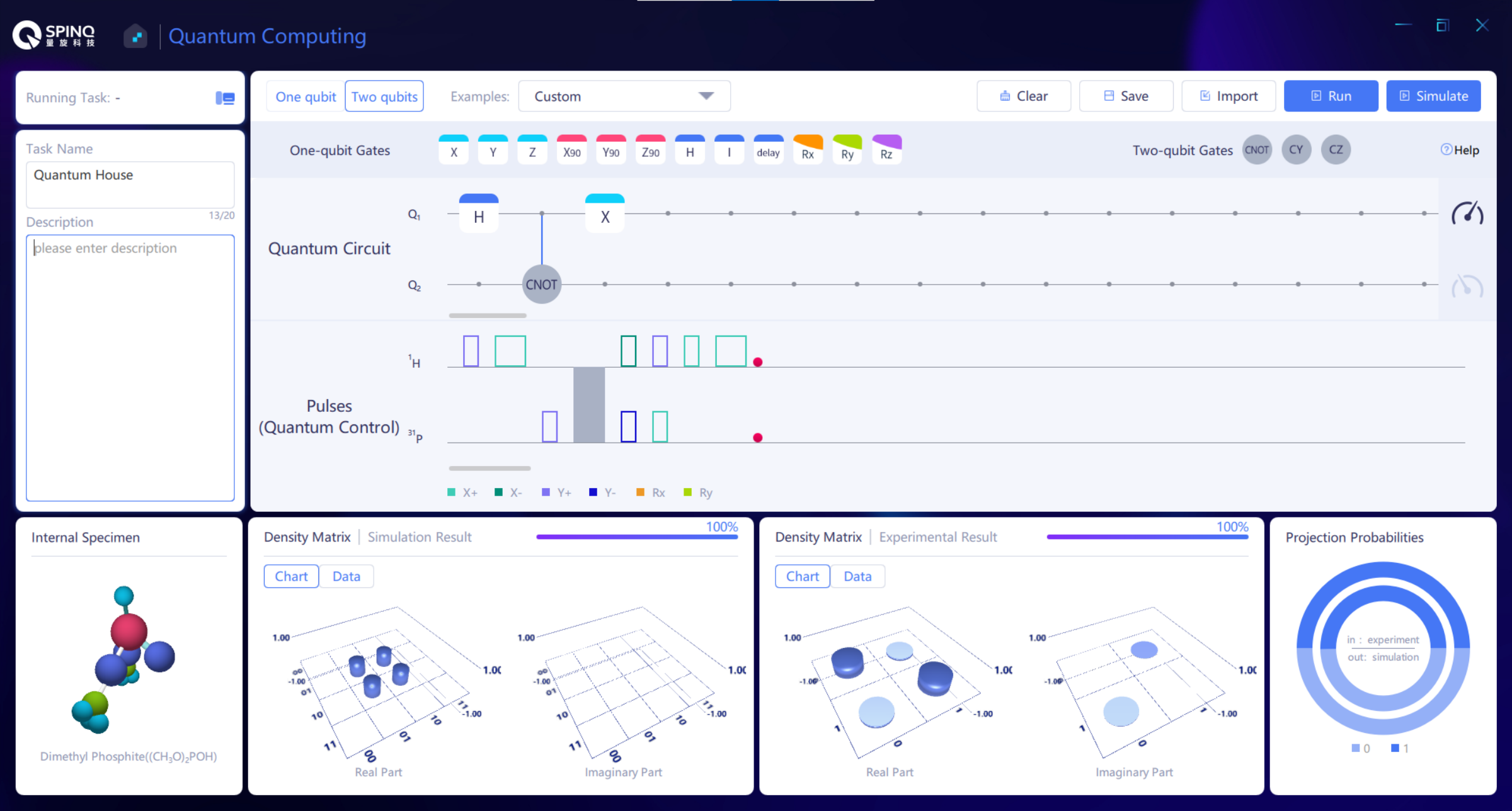}\\
\medskip
\caption{Partial impact of applying a Pauli-$X$ gate on the first qubit of the EPR pair $\rho_{AB}=\frac{1}{2}(\vert 00\rangle+\vert 11\rangle)(\langle 00\vert+\langle 11\vert)$. \textbf{Top screenshot:} the only difference to Fig.~\ref{figepr2q} is that here the "Density Matrix $\vert$ Experimental Result" section shows the noisy density matrix chart of the first qubit only (see Fig.~\ref{figepr1qdata} for numerical format). It is because the measurement icon at the end of "Quantum Circuit" is greyed out for the second qubit. Similarly to Fig.~\ref{figepr2q}, the noisy density matrix is close to the ideal $\rho_A=\mathrm{tr}_B( \rho_{AB})=\frac{I}{2}$, which is the 1-qubit maximally mixed state. On the other hand, the "Density Matrix $\vert$ Simulation Result" section still shows the ideal 2-qubit state $\rho_{AB}$, just like in Fig.~\ref{figepr2q}. \textbf{Bottom screenshot:} after applying an additional Pauli-$X$ gate on the first qubit of the EPR pair, the density matrix chart indicates that the state of the first qubit remains the same (apart from noise), namely $\rho'_A=\mathrm{tr}_B( \rho'_{AB})=\frac{I}{2}$, where $\rho'_{AB}=\frac{1}{2}(\vert 10\rangle+\vert 01\rangle)(\langle 10\vert+\langle 01\vert)$.}\label{figepr1q}
\end{figure*}

Then, the screenshots in Fig.~\ref{figepr1q} show that as opposed to the overall 2-qubit state, the state of the first qubit alone isn't changed (apart from noise) by applying a Pauli-$X$ gate! And this completes the demonstration of the quantum-house effect.

But there is something we shouldn't overlook here. In fact, due to the noise it would be more accurate to say that we only \textit{illustrated} the quantum-house effect, rather than implemented it on the hardware level. This is because, from a cryptographic point of view, the noise characteristics of the quantum device might give Alice enough hints to be able to figure out whether or not Charlie has applied a Pauli-$X$ gate on the first qubit. So the noise always has to be taken into consideration in a realistic situation.

Therefore, we propose an informal definition for the non-ideal case where noise is present, but won't pursue it further in this paper.

\begin{definition}[Noisy quantum-house effect]\label{defnqhe}
The noisy quantum-house effect is the phenomenon when an operation on subsystem $A$ changes the overall state of system $AB$ significantly, while causing only insignificant change to the state of $A$.
\end{definition}

It's like a non-local, immediate butterfly effect, with the twist that in the extreme, noiseless case, even \textit{no change} to subsystem $A$ causes a significant change to system $AB$.

\section{Theory - Part 2}\label{sec4}

Next, we give a characterization of the $\rho_{AB}$ states with which the quantum-house effect can be achieved.

Surprisingly, we'll find that besides non-product states, the quantum-house effect is possible even for some product states $\rho_{AB}=\rho_A\otimes\rho_B$, provided that neither $\rho_A$ nor $\rho_B$ are pure states.

\begin{theorem}[Non-product implies quantum-house]\label{thmnonprod}
The quantum-house effect can be achieved with any non-product state $\rho_{AB}\ne\rho_A\otimes\rho_B$.
\end{theorem}

\begin{proof}
If we swap $A$ with an independently prepared quantum system in state $\rho_A$, then the new overall state for Alice and Bob will be $\rho'_{AB}=\rho_A\otimes\rho_B$, which is a product state, so clearly $\rho'_{AB}\ne\rho_{AB}$.\footnote{Local operations on $A$ never change the quantum state of $B$, due to the no-signaling principle.} From this, we can also see that the state of Alice's subsystem remained $\rho_A$, and thus we have achieved the quantum-house effect.
\end{proof}

\begin{example}\label{exnonprodqhe}
Let $\rho_{AB}=\frac{1}{2}\vert 00\rangle\langle 00\vert+\frac{1}{2}\vert 11\rangle\langle 11\vert$. This is a non-product state, and a straightforward calculation reveals that $\rho_A=\frac{1}{2}\vert 0\rangle\langle 0\vert+\frac{1}{2}\vert 1\rangle\langle 1\vert=\frac{I}{2}$, the 1-qubit maximally mixed state. Now, if Charlie replaces Alice's qubit with an independently prepared qubit in state $\frac{I}{2}$, then the resulting new overall 2-qubit state for Alice and Bob will be $\rho'_{AB}=\frac{1}{2}\frac{I}{2}\otimes\vert 0\rangle\langle 0\vert+\frac{1}{2}\frac{I}{2}\otimes\vert 1\rangle\langle 1\vert=\frac{I}{2}\otimes\frac{I}{2}$, which is a product state and thus different from $\rho_{AB}$. At the same time, the state of Alice's qubit stays $\rho'_A=\rho_A=\frac{I}{2}$.
\end{example}

\begin{theorem}[Non-pure implies quantum-house]\label{thmnontrivialqhe}
The quantum-house effect can be achieved with any product state $\rho_{AB}=\rho_A\otimes\rho_B$ where neither $\rho_A$ nor $\rho_B$ is pure.
\end{theorem}

\begin{proof}
Let $\sigma_{A'B'}$ be  a non-product state of some bipartite system $A'B'$ with $\sigma_{A'}=\rho_A$ and $\sigma_{B'}=\rho_B$. Such a state can always be produced, because both $\rho_A$ and $\rho_B$ have support containing more than one element, and thus they can be made classically correlated with each other. We give $B'$ to Bob (i.e. $B=B'$), and keep $A'$ for ourselves. Additionally, we independently prepare another system $A$ in state $\rho_A$, and give that to Alice. Now, the overall state of the system possessed by Alice and Bob is $\rho_{AB}=\rho_A\otimes\rho_B$. Then, if we swap Alice's subsystem $A$ with the $A'$ we kept before, the overall state for Alice and Bob changes to $\rho'_{AB}=\sigma_{A'B'}$, which is different from $\rho_{AB}$. But since the state of Alice's subsystem remains $\rho'_A=\sigma_{A'}=\rho_A$, we have achieved the quantum-house effect.
\end{proof}

An important difference between Theorem~\ref{thmnonprod} and Theorem~\ref{thmnontrivialqhe} is that the proof of the latter requires that there is side-information available which is correlated with Bob's subsystem,\footnote{Thus, the swapping operation in the proof of Theorem~\ref{thmnontrivialqhe} is "local" only in a \textit{geographical} sense, in that it is performed inside Alice's lab.} while for the former theorem it's enough if we just know $\rho_{A}$.

\begin{example}\label{exprodqhe}
Let Charlie first prepare $\sigma_{A'B'}=\frac{1}{2}\vert 00\rangle\langle 00\vert+\frac{1}{2}\vert 11\rangle\langle 11\vert$, which is a non-product state with $\sigma_{A'}=\sigma_{B'}=\frac{I}{2}$. Charlie gives the second qubit to Bob, and keeps the first for himself. Then, he prepares a new qubit, independently in state $\frac{I}{2}$, and gives that to Alice. Thus, for Alice and Bob the overall 2-qubit state is $\rho_{AB}=\frac{I}{2}\otimes\frac{I}{2}$. Now, if Charlie swaps Alice's qubit with the one he kept before, the overall 2-qubit state for Alice and Bob will change to $\rho'_{AB}=\sigma_{A'B'}$, which is different from $\rho_{AB}$. However, the state of Alice's qubit remains the same: $\rho'_A=\sigma_{A'}=\frac{I}{2}$.
\end{example}

Finally, it's easy to see that with the previous theorem we've reached the limit:

\begin{theorem}[Pure implies no quantum-house]\label{thmtrivialnoqhe}
The quantum-house effect cannot be achieved with any product state $\rho_{AB}=\rho_A\otimes\rho_B$ where either $\rho_A$ or $\rho_B$ is pure.
\end{theorem}

\section{Discussion}\label{sec5}

We introduced the quantum-house effect, a non-local phenomenon which can be exhibited even with bipartite product states, provided that neither subsystem is in a pure state. The effect was demonstrated (with some inevitable noise) on the SpinQ Gemini 2-qubit liquid-state NMR desktop quantum computer.

Since the quantum-house effect can be achieved also with non-entangled states, it extends the notion of quantum nonlocality to a wider range of bipartite quantum states than entanglement would allow, meaning that separable quantum states can already behave in a counter-intuitive way in this regard, and thus exhibit non-classicality.

To go beyond the quantum-house effect, we suggest that one way to characterize the point where quantumness departs from classicality is via \textit{quantum detachment}, a principle which roughly means that relevant information about the state of a physical system\footnote{Information that would influence what outcomes an experimenter may expect when physically interacting with the system.} is kept separate from the system itself. In the examples throughout this paper, Charlie successfully utilized that state-relevant information was detached from Alice's subsystem $A$, thereby rendering Alice's task of figuring out things on her own impossible.\footnote{The idea of quantum detachment can be conveyed as follows: when the locally unavailable information is somewhere else, we have a mixed state; and when it's nowhere else, we have superposition. The latter case can be considered as the ultimate quantum detachment, because the missing information (e.g. about the outcome of a future measurement) doesn't even exist in the universe.}

We can contrast quantum detachment with the classical world where no relevant information about the state of a physical system can be detached, as all of it can be found out locally in the lab, in principle. Put differently, in the classical world a physical system contains all the relevant information about itself.

As for future work, the quantum-house effect might be used to create a protocol by which Charlie could securely cast a "yes/no" vote, locally in Alice's lab, without using entangled quantum states.

\backmatter

\begin{figure*}[p]
\centering
\includegraphics[width=0.98\linewidth]{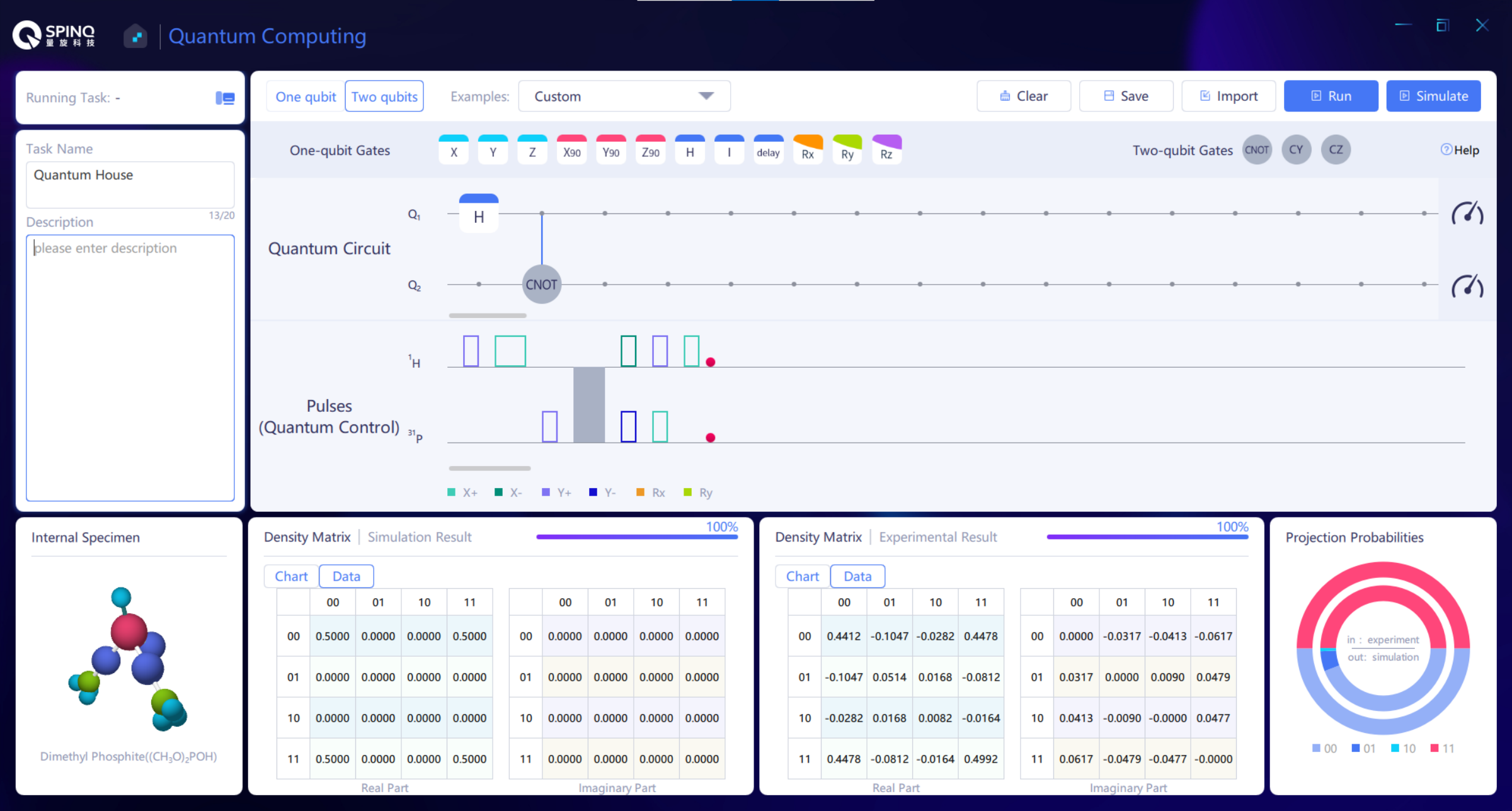}\\
\bigskip
\includegraphics[width=0.98\linewidth]{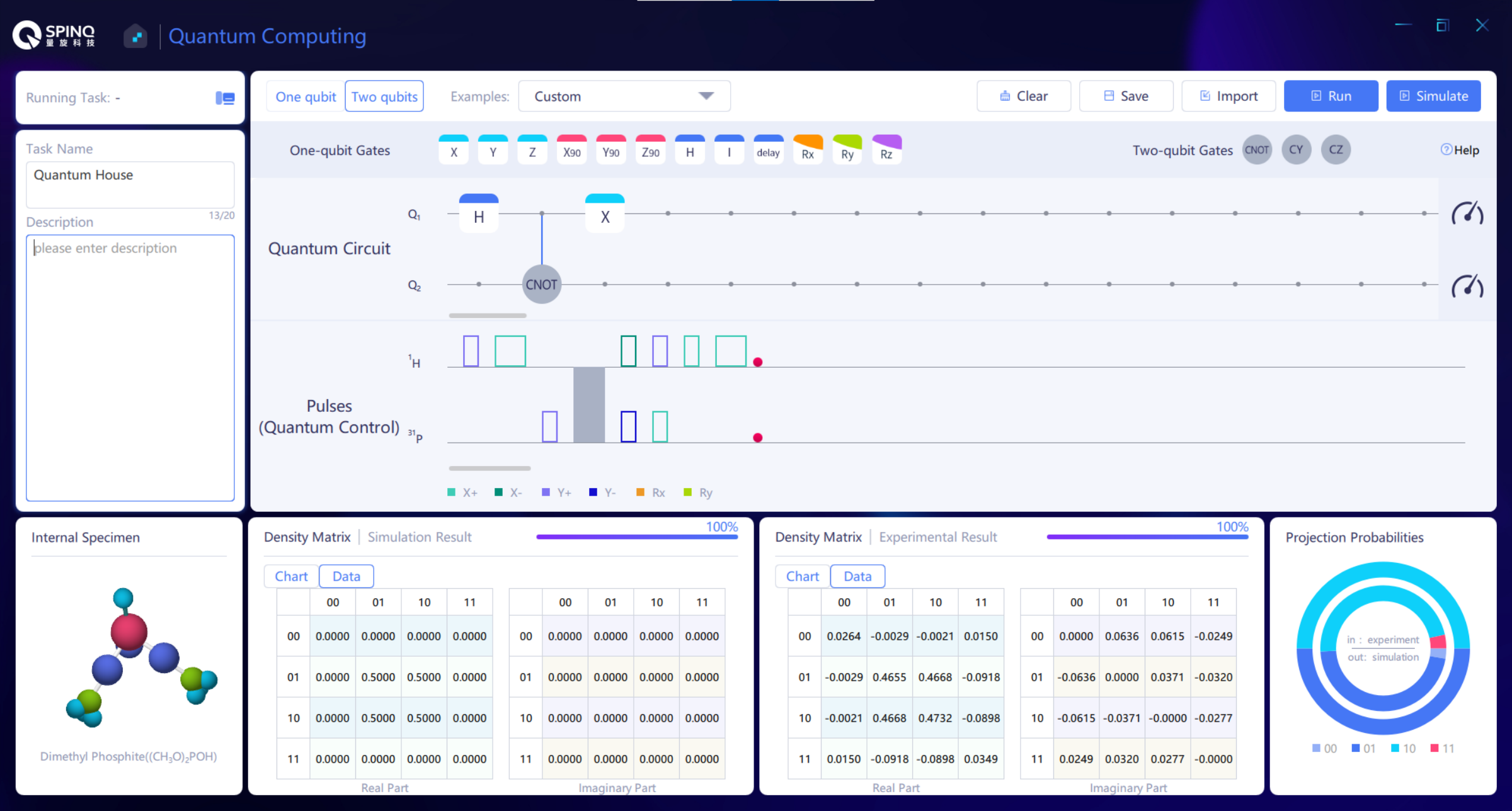}\\
\medskip
\caption{The "Data" view of Fig.~\ref{figepr2q}, showing the numerical density matrices.}\label{figepr2qdata}
\end{figure*}

\begin{figure*}[p]
\centering
\includegraphics[width=0.98\linewidth]{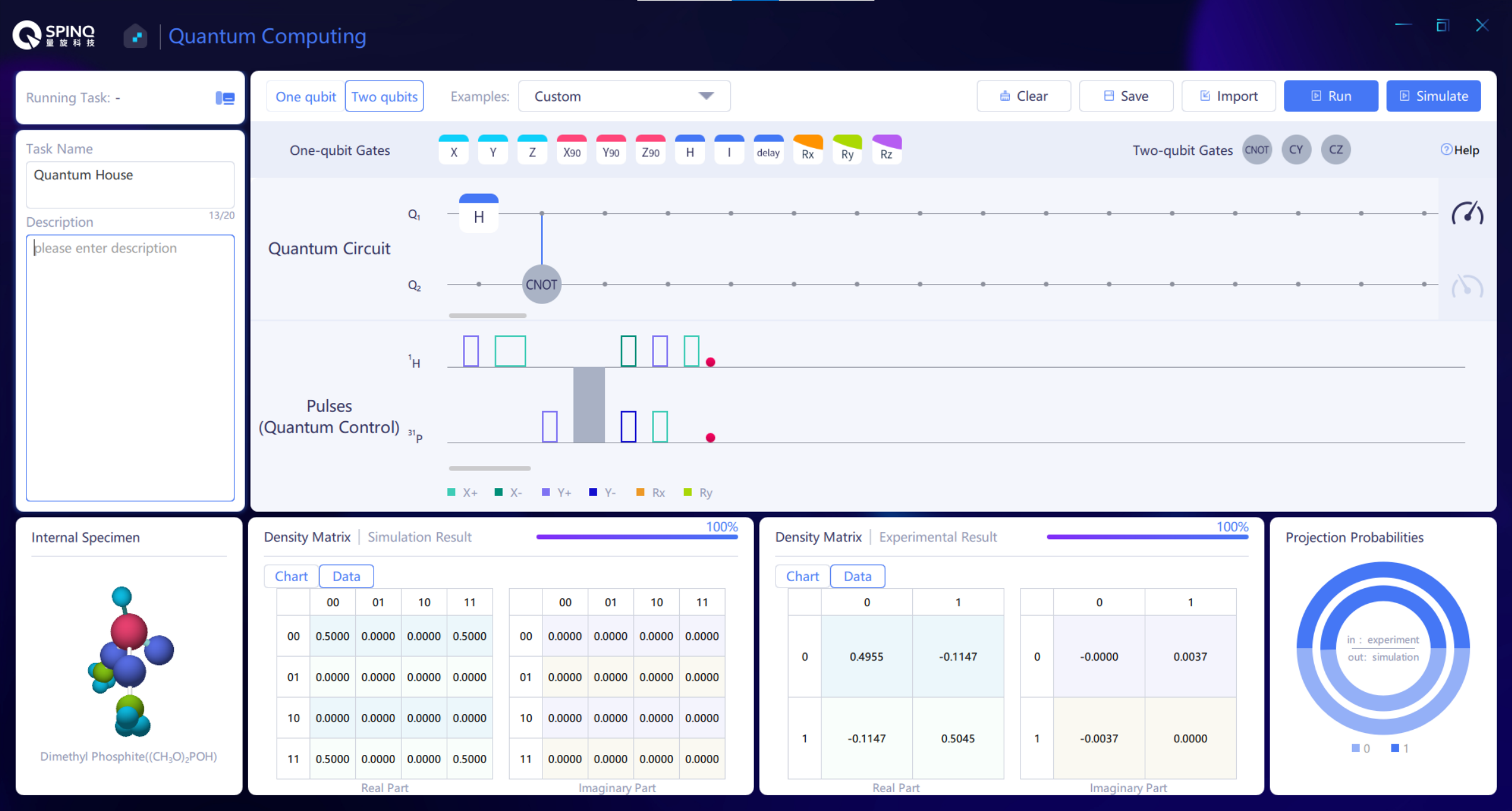}\\
\bigskip
\includegraphics[width=0.98\linewidth]{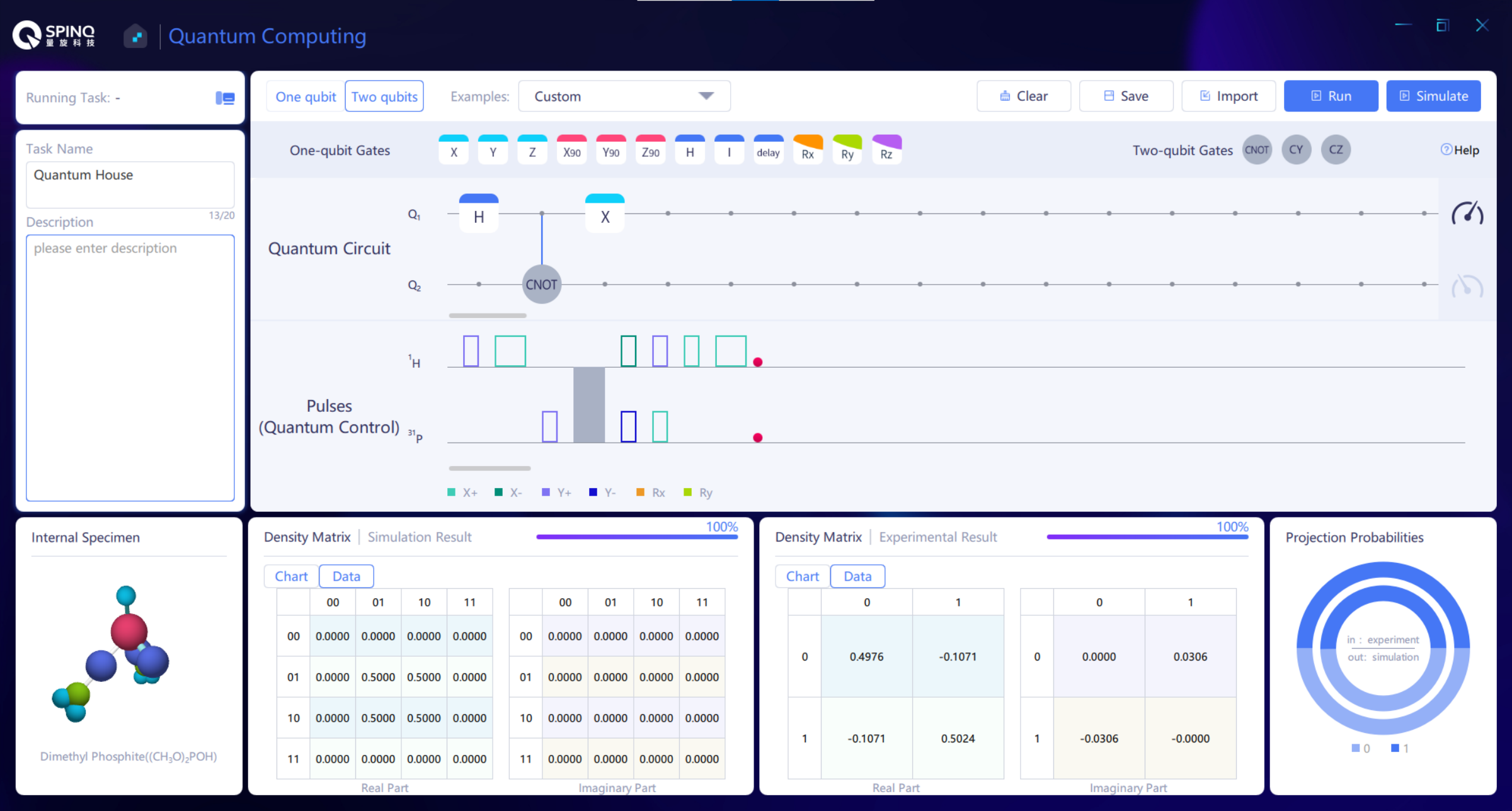}\\
\medskip
\caption{The "Data" view of Fig.~\ref{figepr1q}, showing the numerical density matrices.}\label{figepr1qdata}
\end{figure*}

\end{document}